\documentclass[cernpreprint, atlasdraft=false, texlive=2016, UKenglish, texmf]{atlasdoc}
 
\usepackage{atlaspackage}
\usepackage{atlasbiblatex}
 
\usepackage{atlasphysics}
\usepackage{footmisc}
 
\graphicspath{{logos/}{figures/}}
 
\usepackage{ANA-SIMU-2020-01-PAPER-defs}

\AtlasTitle{Emulating the impact of additional proton--proton interactions in the ATLAS simulation by
presampling sets of inelastic Monte Carlo events}

\AtlasAbstract{
ABSTRACT \\
 
The accurate simulation of additional interactions at the ATLAS experiment for the analysis of
proton--proton collisions delivered by the Large Hadron Collider presents a significant
challenge to the computing resources. During the LHC Run~2 (2015--2018) there were up to 70 inelastic
interactions per bunch crossing, which need to be accounted for in Monte Carlo (MC) production. In this
document, a new method to account for these additional interactions in the simulation chain is
described. Instead of sampling the inelastic interactions and adding their energy deposits to a hard-scatter
interaction one-by-one, the inelastic interactions are presampled, independent of the hard scatter,
and stored as combined events. Consequently, for each hard-scatter interaction only one such presampled
event needs to be added as part of the simulation chain.
For the Run~2 simulation chain, with an average of 35 interactions per bunch crossing, this new method
provides a substantial reduction in MC production CPU needs of around 20\%, while reproducing the properties of the  
reconstructed quantities relevant for physics analyses with good accuracy.
}
 
\author{The ATLAS Collaboration}
 
\AtlasRefCode{SIMU-2020-01}
 
\PreprintIdNumber{CERN-EP-2021-007}

\AtlasJournalRef{Comput Softw Big Sci 6, 3 (2022)}
\AtlasDOI{10.1007/s41781-021-00062-2}
\hypersetup{pdftitle={Emulating the impact of additional proton-proton interactions in the ATLAS simulation by presampling sets of inelastic Monte Carlo events},pdfauthor={The ATLAS Collaboration}}
 
\begin{document}
 
\maketitle
 
\tableofcontents

\section{Introduction}
The excellent performance of the Large Hadron Collider (LHC) creates a challenging environment for
the ATLAS and CMS experiments. In addition to the hard-scatter proton--proton (\( pp \)) interaction
which is of interest for a given physics analysis, a large number of inelastic proton--proton
collisions occur simultaneously.
These are collectively known as \textit{\pileup}. The mean number of these inelastic \( pp \) interactions per bunch crossing, \actualmu, also known as the \pileup parameter, characterises the instantaneous luminosity at any given time\footnote{Hereafter, the approximation that \actualmu is the same for all colliding bunch pairs is made.}.
 
For physics analyses, \pileup is conceptually similar to a noise contribution that needs to be
accounted for as it is unrelated to the hard--scatter event that is of interest for the analysis. Since nearly all analyses rely on Monte Carlo (MC) simulation to predict the detector response to the physics process, it is crucial that the \pileup is modelled correctly as part of that
simulation. The goal of the ATLAS MC simulation chain is to accurately reproduce the \pileup such that it can be accounted for in physics analyses.

\begin{figure}[htbp]
\centering
\includegraphics[width=0.55\textwidth]{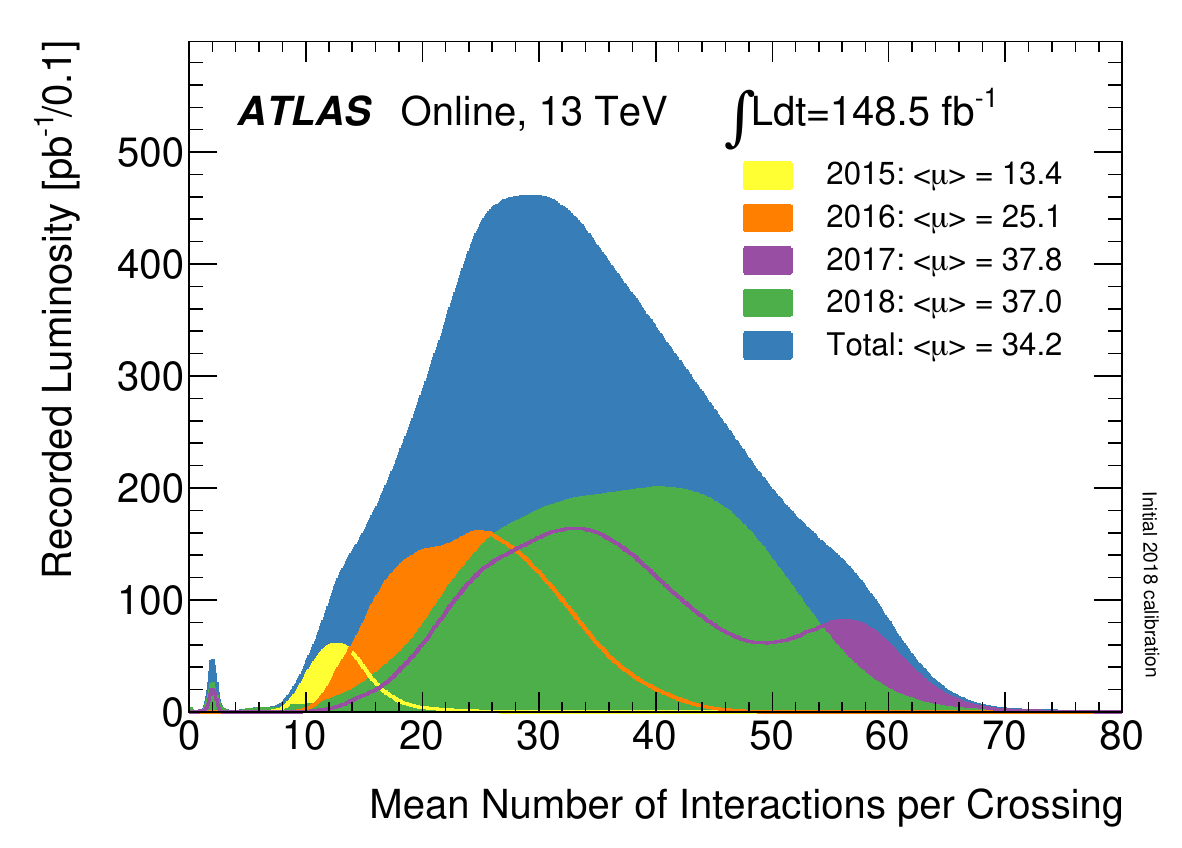}
\caption{The \actualmu distribution observed for the ATLAS Run 2 data, for each year (2015--2018) separately and for the sum of all years~\cite{lumi}.}
\label{fig:mu}
\end{figure}

Within ATLAS, the \pileup is emulated by overlaying soft inelastic \( pp \) interactions, in the
following called minimum-bias interactions, generated with an MC generator,
normally \PYTHIA~\cite{Sjostrand:2007gs}, according to the \pileup profile for a given data-taking
period. \Fig{\ref{fig:mu}} shows the \actualmu distribution for each year during \RunTwo  (2015--2018) and the sum of
all years. The mean value is \( 34.2 \) but the distribution is broad and generally covers values
between 10 and 70. The small peak at \( \mu\sim 2 \) arises from special running periods with rather low
luminosity. At the High Luminosity LHC (HL-LHC) \actualmu is expected to increase to about 200~\cite{1402887}. The inelastic interactions include non-diffractive and diffractive interactions based on the  Donnachie--Landshoff~\cite{Donnachie_1992} model for the cross sections of the individual processes.
 
The \simc for MC events contains several steps, starting from the generation of the interactions
with an MC generator (e.g.\ \PYTHIA, \SHERPA~\cite{Gleisberg:2008ta}).
The interactions of the generated particles with the ATLAS detector are simulated using a \(\GEANT\)-based~\cite{Agostinelli:2002hh} simulation
framework~\cite{SOFT-2010-01}. This is performed separately for the hard-scatter interactions of
interest and a large number of minimum-bias interactions. Next, the readout of the detector is
emulated via a process known as \textit{digitisation}, which takes into account both the
hard-scatter and any overlapping minimum-bias interactions. In this article, two methods of
performing the digitisation are compared. The goal of the new method, described below, is to
reduce the computing resources required by creating a large set of \pileup events only once for an MC production campaign and then reusing these events for different hard-scatter events. A similar method has been explored by the CMS Collaboration~\cite{CMSOverlay}.
 
In the first method, referred to as \textit{\standardmethod} hereafter, the hard-scatter interaction and the desired number of minimum-bias interactions are read in simultaneously during the digitisation step and the energy deposits made by particles are added for each detector element.  Then the detector readout is emulated to convert these into digital signals, which are finally used in the event reconstruction. This method creates the \pileup on demand for each hard-scatter event, and
has been used up to now for all ATLAS publications based on \pp collisions. In the second (and
new) method, referred to as \textit{\mcoverlay} hereafter, this same procedure is followed but for
the set of minimum-bias interactions alone, without the hard-scatter interaction. The resulting
presampled events are written out and stored.  Then,
during the digitisation of a given hard-scatter interaction, a single presampled event is picked
and its signal added to that of the hard-scatter interaction for each readout channel.
This combined event is then input to the event reconstruction. In contrast to the first method, the same
presampled \pileup event can be used for several hard-scatter interactions. For both methods, the \actualmu value to be used is
sampled randomly from the data \actualmu distribution, such that the ensemble of many events follows the \actualmu distribution of the data.
 
If the detector signals were read out without any information loss, the two methods would give
identical results. However, in reality some information loss occurs due to readout thresholds
applied or custom compression algorithms designed to reduce the data volume. This can lead to
differences in the reconstructed quantities used in physics analyses. While in most cases for ATLAS
these differences were found to be negligible, in some cases corrections were derived to reduce the
impact on physics analyses, as is discussed in \Sectrange{\ref{sec:indet}}{\ref{sec:trig}}.
 
Within the ATLAS Collaboration, a significant validation effort took place to ensure that
this \mcoverlay \simc reproduces the results from the \standardmethod \simc accurately, so that there is no impact on physics analyses whether one or the other is used.  To this end, thousands of
distributions were compared between the \mcoverlayandstd \simcs. In this article, a
representative subset of relevant distributions is shown.
Only comparisons between the two methods are shown in this article; detailed comparisons of data with simulation can be found in various
performance papers, see e.g.\
\Refns{\cite{PERF-2015-01,EGAM-2018-01,PERF-2015-10,JETM-2018-05,FTAG-2018-01,ATLAS-CONF-2017-029}}.
 
The motivation for using the \mcoverlay \simc in the future is that it uses significantly less CPU time
than the \standardmethod \simc. As is discussed in \Refn{\cite{CompCDR}}, savings in CPU, memory and
disk space requirements are pivotal for the future running of the ATLAS experiment.  Additionally,
the \mcoverlay \simc can also be seen as a step towards using minimum-bias data, instead of
presampled simulated events, for emulating the \pileup, which could potentially improve the accuracy of the
modelling of the \pileup interactions.  However, the \pileup emulation with data is not yet
validated and not the subject of this article.

The article is organised as follows. A description of the ATLAS detector is given in
\Sect{\ref{sec:det}}, highlighting the aspects that are most relevant for the \pileup
emulation. \Sect{\ref{sec:overview}} describes both the \standardandover \simc,
and \Sect{\ref{sec:perf}} compares their CPU and memory performances. In \Sectrange{\ref{sec:indet}}{\ref{sec:trig}} the challenges in the inner detector,
calorimeters, muon system and trigger are described and comparisons of the impact of the old and new methods are shown. For these comparisons a variety of different MC samples are used, based on what is most appropriate for the validation of a given object within the particular detector subsystem.
 
For all studies presented in this article, unless otherwise stated, the distribution of the average number of events per bunch crossing follows the distribution observed in the ATLAS data in 2017, with an average \( \mu \) value of \(37.8\) (see \Fig{\ref{fig:mu}}). The ATLAS detector configuration corresponds to that of \RunTwo.  As the detector configuration evolves in the future, the new \mcoverlay method will need to be validated for those new detector elements.
 
 
\section{ATLAS detector}
\label{sec:det}
\newcommand{\AtlasCoordFootnote}{
ATLAS uses a right-handed coordinate system with its origin at the nominal interaction point (IP) in
the centre of the detector and the \(z\)-axis along the beam pipe.  The \(x\)-axis points from the IP to
the centre of the LHC ring, and the \(y\)-axis points upwards.  Cylindrical coordinates \((r,\phi)\) are
used in the transverse plane, \( \phi \) being the azimuthal angle around the \(z\)-axis.  The
pseudorapidity is defined in terms of the polar angle \( \theta \) as \(\eta = -\ln \tan(\theta/2)\).
Angular distance is measured in units of \(\Delta R \equiv \sqrt{{(\Delta\eta)}^{2} +
{(\Delta\phi)}^{2}}\).}
 
The ATLAS detector~\cite{PERF-2007-01}
at the LHC covers nearly the entire solid angle around the
collision point.  It consists of an inner tracking detector surrounded by a thin superconducting
solenoid, electromagnetic and hadronic calorimeters, and a muon spectrometer incorporating three
large superconducting toroidal magnets.  A two-level trigger system is used to select interesting
events~\cite{TRIG-2016-01}.  The first-level (L1) trigger is implemented in hardware and uses a subset
of detector information to reduce the event rate from \SI{40}{MHz} to \SI{100}{kHz}.  This is
followed by a software-based high-level trigger (HLT) which reduces the event rate to an
average of \SI{1}{kHz}.
 
At the LHC, typically 2400 bunches from each of the two proton beams cross each other at the ATLAS interaction point per beam revolution,
with one bunch crossing (BC) taking place every \SI{25}{ns}. In each BC several \pp interactions may occur.
Whenever an L1 trigger signal is received for a given BC the entire detector is read out and processed in the HLT to decide
whether the event is stored for further analysis.
 
The inner detector (ID) is immersed in a \SI{2}{\tesla} axial magnetic field and provides
charged-particle tracking in the pseudorapidity\footnote{\AtlasCoordFootnote} range \(\abseta < 2.5\). The
high-granularity silicon pixel detector (Pixel), including an insertable B-layer
(IBL)~\cite{ATLAS-TDR-19,PIX-2018-001} added in 2014 as a new innermost layer, covers the vertex region and
typically provides four measurements per track, the first hit normally being in the innermost layer.
It is followed by the silicon microstrip tracker (SCT) which usually provides four two-dimensional
measurement points per track.  These silicon detectors are complemented by a straw tracker
(transition radiation tracker, TRT), which enables radially extended track reconstruction with an
average of \(\sim 30\) hits per track up to \(\abseta = 2.0\).  Additionally, the transition radiation
capability provides separation power between electrons and charged pions.
 
The calorimeter system covers the pseudorapidity range \(\abseta < 4.9\).  Within the region \(\abseta <
3.2\), electromagnetic (EM) calorimetry is provided by barrel (EMB) and endcap (EMEC) high-granularity
lead/liquid-argon (LAr) electromagnetic calorimeters, with an additional thin LAr presampler
covering \(\abseta < 1.8\) to correct for energy loss in material upstream of the calorimeters.
Hadronic calorimetry is provided by the steel/scintillator-tile (Tile) calorimeter, segmented into three
barrel structures within \(\abseta < 1.7\), and two copper/LAr hadronic endcap calorimeters (HEC).  The
solid angle coverage is completed with forward copper/LAr and tungsten/LAr calorimeter (FCAL)
modules optimised for electromagnetic and hadronic measurements, respectively.
 
The muon spectrometer (MS) comprises separate trigger and high-precision tracking chambers measuring
the deflection of muons in a toroidal magnetic field generated by the superconducting air-core magnets.  The
field integral of the toroids ranges between \num{2.0} and \SI{6.0}{\tesla\metre} across most of the
detector.  A set of precision chambers covers the region \(\abseta < 2.7\) with three stations of
monitored drift tubes (MDTs), complemented by cathode strip chambers (CSCs) in the forward region, where the
background is highest.  The muon trigger system covers the range \(\abseta < 2.4\) with resistive plate
chambers (RPCs) in the barrel, and thin gap chambers (TGCs) in the endcap regions.
 
The integration times of the different subdetectors vary significantly, mostly due to the charge
drift times depending on the material and geometry of the respective detector system. In most cases,
the integration time exceeds \SI{25}{ns}, i.e.\ the time between two BCs.  In such cases, the signal from
events that occurred in previous BCs contaminates the signal in the triggered BC.\@ This is often
referred to as out-of-time \pileup and needs to be considered for the simulation, in
addition to the in-time \pileup which accounts for signals generated by interactions occurring
inside the BC corresponding to the hard-scatter event.
 
\begin{figure}[htbp]
\centering \includegraphics[width=0.65\textwidth]{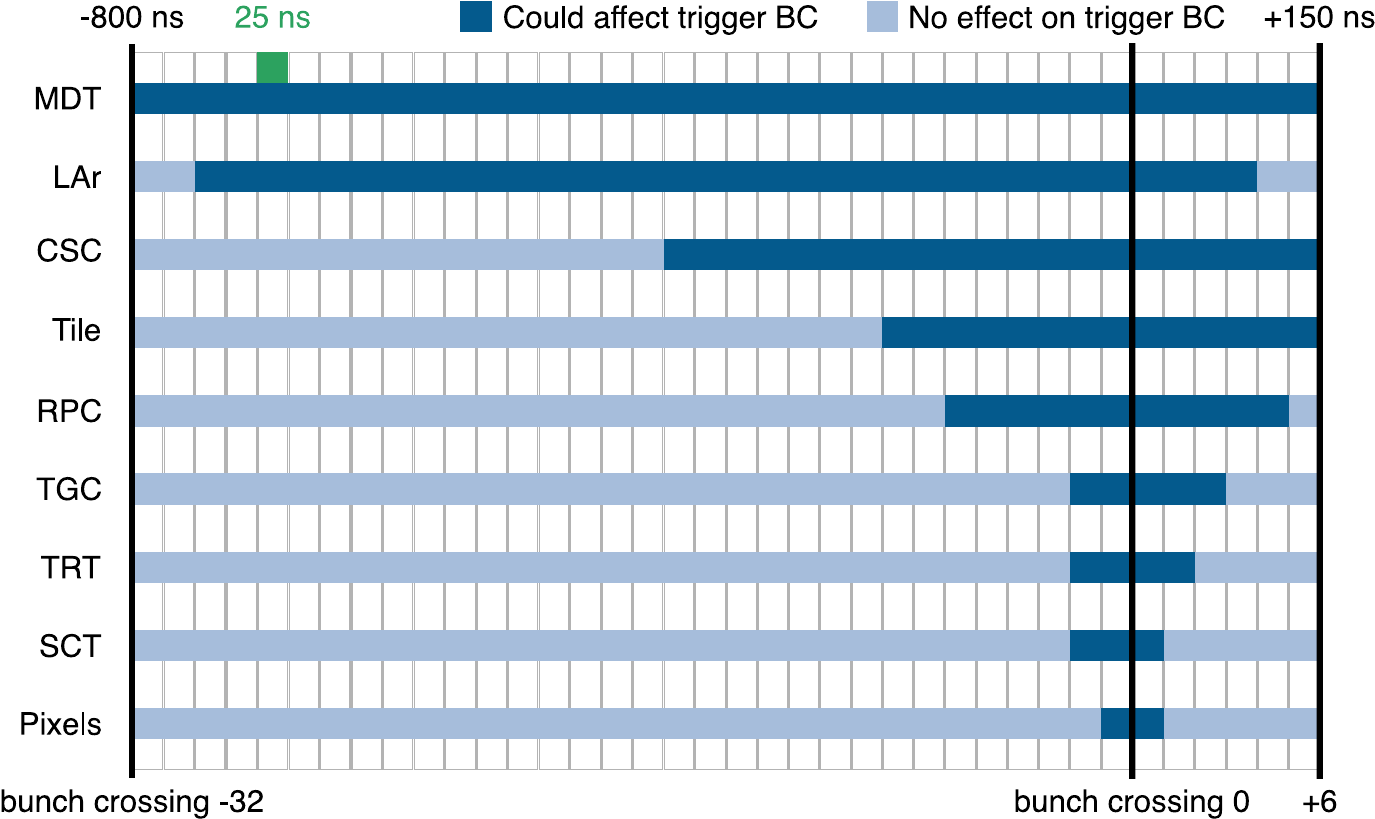} \caption{The time windows
considered for the simulation of each subdetector. The dark blue BCs are those where a signal in that
BC can contaminate the signal in the triggered BC (i.e.\ BC 0), while the light blue coloured BCs cannot affect the
triggered BC.}
\label{fig:readout}
\end{figure}

\Fig{\ref{fig:readout}} shows the readout windows considered for the simulation of each of the
detector systems. The MDTs have the longest integration time, \SI{750}{ns}, with 32 BCs prior to the
trigger and 6 BCs after the trigger being considered. For the LAr calorimeter it is only slightly shorter.
For the inner detector (Pixel, SCT and TRT) the integration time is much shorter, and only the 1--2 BCs
before and after the trigger need to be considered.
 
\section{Overview of simulation chain}
\label{sec:overview}
As is described above, the ATLAS simulation chain~\cite{SOFT-2010-01}, used to produce MC samples to be
used in physics and performance studies, is divided into three steps: generation of the event and
immediate decays, particle tracking and physics interactions in the detector, based
on \(\GEANT\)~(G4), and digitisation of the energy deposited in the sensitive regions of the
detector into voltages and currents to emulate the readout of the ATLAS detector. This simulation
chain is integrated into the ATLAS software framework, \textsc{Athena}~\cite{ATLAS-TDR-17}.
Finally, a series of reconstruction algorithms is applied in the same way as for the data, where
final physics objects such as jets, muons and electrons are reconstructed~\cite{PERF-2007-01}.  Each
step can be run as an individual task, but in order to save disk space the digitisation step is
usually performed in the same task as the reconstruction step, such that the intermediate output
format from the digitisation step only needs to be stored locally on the computing node and can be
discarded after the reconstruction step is finished.
 
The G4 simulation step is run by itself and, since it is independent of the detector readout
configuration, the trigger and the \pileup, it is often run significantly earlier than the
digitisation and reconstruction, which depend on these aspects. The G4 simulation is the most CPU
intensive and thus it is desirable to run this as rarely as possible.
 
The ATLAS digitisation software converts the energy deposits (HITS) produced by the G4 simulation in the sensitive elements
into detector response objects, known as \textit{digits}. A digit is produced when the voltage or current of a particular
readout channel rises above a preconfigured threshold within a particular time window.  Some of the
subdetectors read out just the triggered BC, while others read out several bunch crossings,
creating digits for each.  For each digit, some subdetectors (e.g.\ SCT) record only the fact that a given
threshold has been exceeded, while others (e.g.\ Pixel or LAr) also retain information related to the
amplitude. The digits of each subdetector are written out as Raw Data Objects (RDOs), which contain information about the readout channel identifier and the raw data that is sent from the detector front-end electronics.
 
\begin{figure}[htpb]
\centering
\includegraphics[width=\textwidth]{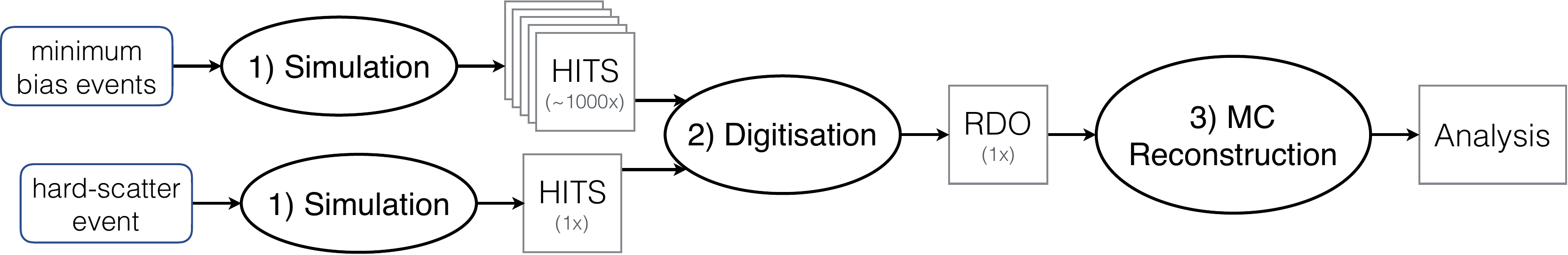}
\caption{Current workflow diagram from simulation to physics analysis.
The oval steps represent an action while the boxes represent data files of a
given format. The final box is the reconstructed data in analysis
format.}\label{fig:digi_workflow}
\end{figure}
 
For any given hard-scatter interaction, the additional \pileup interactions must be included in a
realistic model of the detector response. For this purpose, minimum-bias events are generated using
the \PYTHIA event generator with the NNPDF2.3LO~\cite{Ball:2012cx} parton distribution function and the  
A3~\cite{ATL-PHYS-PUB-2016-017} set of tuned parameters, then simulated and stored in separate
files.
In order to avoid potential issues due to low statistics of relatively hard minimum-bias events, containing objects with high transverse momentum (\( \pt \)), the sampled events are in fact split into two distinct equal-sized samples: a high-\( \pt \) sample composed of events containing jets or photons with \(\pt > \SI{35}{\GeV}\) and a low-\( \pt \) sample composed of the remaining events.  Events are then selected from each sample based on their relative cross-section to avoid duplicating distinctive hard events, which may give rise to visible features during analysis.

In the current \standardmethod \simc, the simulation files of both the hard-scatter event and the desired number of  minimum-bias events are read in concurrently at the digitisation step and the HITS are combined.  For each hard-scatter event a value of \actualmu is assigned by randomly sampling the \actualmu distribution corresponding to the relevant data-taking period.  Most subdetector responses are affected by interactions from neighbouring bunch crossings: as is shown in \Fig{\ref{fig:readout}}, up to 32 BCs before and 6 BCs after the triggering BC may contribute signal to the trigger BC.\@  For the
average \( \mu \) value of \(37.8\) during 2017 data taking, this implies that simulating the impact
of \pileup on any given hard-scatter event requires approximately \( (32+1+6) \times 38 = 1482 \) minimum-bias events on average to be selected at random (from the simulated event files) and processed as part of the digitisation step.  Each of these bunch crossings is taken to have the same value of \actualmu as the trigger bunch crossing\footnote{In data there are variations between adjacent bunches of order 10\%~\cite{DAPR-2013-01} but this is not emulated in the MC simulation.}. The number of minimum-bias events (\(N\)) to include for each bunch crossing is drawn at random from a Poisson distribution with a mean of the \actualmu value for that bunch crossing.
After the energy deposits in the trigger BC due to all contributing BCs have been combined, the detector
response is emulated.  This workflow is illustrated in \Fig{\ref{fig:digi_workflow}}.
 
\begin{figure}[htbp]
\centering
\includegraphics[width=0.99\textwidth]{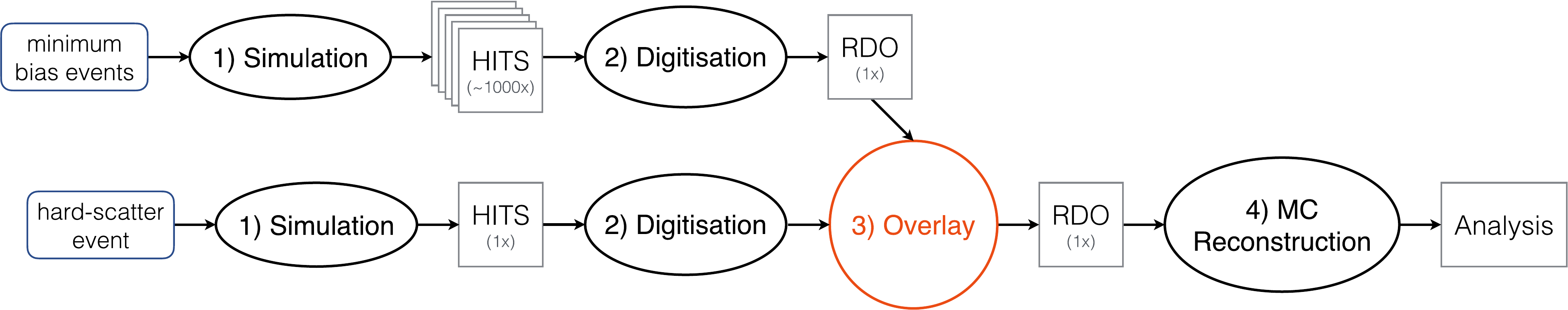}
\caption{The \mcoverlay workflow schema. The oval steps represent an action while the boxes
represent data files of a given format. The final box is the reconstructed data in analysis
format.}
\label{fig:overlay_workflow}
\end{figure}

The new  \mcoverlay \simc is illustrated in \Fig{\ref{fig:overlay_workflow}}.
Rather than digitising the minimum-bias interactions each time a hard-scatter event is produced, a large sample of \pileup events is produced by pre-combining the simulated \pileup interactions, according to the \actualmu distribution of the data campaign, during a separate digitisation step, termed
presampling\footnote{For the calorimeters and the SCT, the digitised output stored in the presampled
events is amended, so that the \mcoverlay can be applied accurately.}. Here, the sampling is done exactly as for the \standardmethod, the only difference being that there is no hard-scatter event. These presampled \pileup events are written out in RDO format as \textit{\pileup RDO datasets} and typically contain several
million events.  Each simulated hard-scatter interaction is then digitised and combined with an
event sampled from these \pileup datasets (step 3 in \Fig{\ref{fig:overlay_workflow}}, called \textit{overlay}).   Here, instead of HITS for each channel, the signals of the RDO or digit (depending on the subdetector) in the hard-scatter event and the presampled event are overlaid.  To avoid double-counting, instrumental noise associated with the detector electronics is included solely in the presampled RDOs and not included in the subsequent hard-scatter digitisation.
Since the digitisation, presampling and reconstruction steps are typically combined into a single task in the production workflow, the output is written locally to an RDO file that is then input to the
reconstruction software; this local RDO file is subsequently discarded. The \pileup RDO datasets
necessary for a given digitisation task are about five times smaller than the many minimum-bias HITS required in the \standardmethod \simc.

The main benefit of the \mcoverlay \simc is that the CPU and I/O requirements of the digitisation are
significantly lower and have a much smaller dependence on \actualmu, as is discussed in \Sect{\ref{sec:perf}}.
However, if a threshold or compression has been applied to the signal when writing the RDO/digit, this results in some loss of information and thereby could reduce the accuracy of the simulation when using
the \mcoverlay method, as is discussed in \Sectrange{\ref{sec:indet}}{\ref{sec:trig}}.
 
For all the comparisons shown in these sections the hard-scatter events are identical for the two methods but the \pileup events are different. This makes the estimation of the uncertainties difficult as the hard-scatter is fully correlated while the \pileup is not. As most of the quantities are selected to be sensitive to \pileup, the uncertainties are calculated assuming the two samples are uncorrelated but in some distributions this leads to an overestimate of the uncertainties, e.g.\ in the reconstruction efficiencies of tracks and leptons and in the trigger efficiencies.

\section{Computing performance comparison}
\label{sec:perf}
 
In this section the performances of the two simulation chains are compared in terms of
CPU time, memory usage and I/O. The validation in terms of physics performance is presented in subsequent sections.
 
The main computing performance benefit of the \mcoverlay \simc stems from the fact a \pileup
dataset is only created once per MC production campaign, and then the individual events within that dataset are used for multiple hard-scatter MC samples, as opposed to being created on demand independently for each MC sample. An MC production campaign happens typically once per data-taking period and comprises billions (B) of hard-scatter events and thousands of individual samples.
A sample is defined as a set of MC
events generated using the same input parameters, e.g.\ a sample of \ttbar events produced by a
certain MC generator with a given set of input parameters. The same presampled \pileup event can
thus be overlaid on many different hard-scatter events from different MC samples. In doing so, care
needs to be taken to ensure that no undesirable effects on physics analyses occur due to reusing the
same \pileup events, as is discussed below.
 
In ATLAS, typically 70\% of the CPU resources are devoted to MC production via the \simc; the remainder is used for data processing and user analyses. At present, with the \RunTwo \pileup profile, the \simc CPU usage is broken down into about 15\% for event generation, 45\% for G4 simulation, 20\% for digitisation and 20\% for other tasks (reconstruction, trigger, event writing).
The \mcoverlay scheme decreases the digitisation time to a negligible level by reusing the \pileup events more efficiently and thus reduces the overall CPU resources required for MC production by about 20\%, as is discussed below.
 
\begin{figure}[!htb]
\centering
\includegraphics[width=0.55\textwidth]{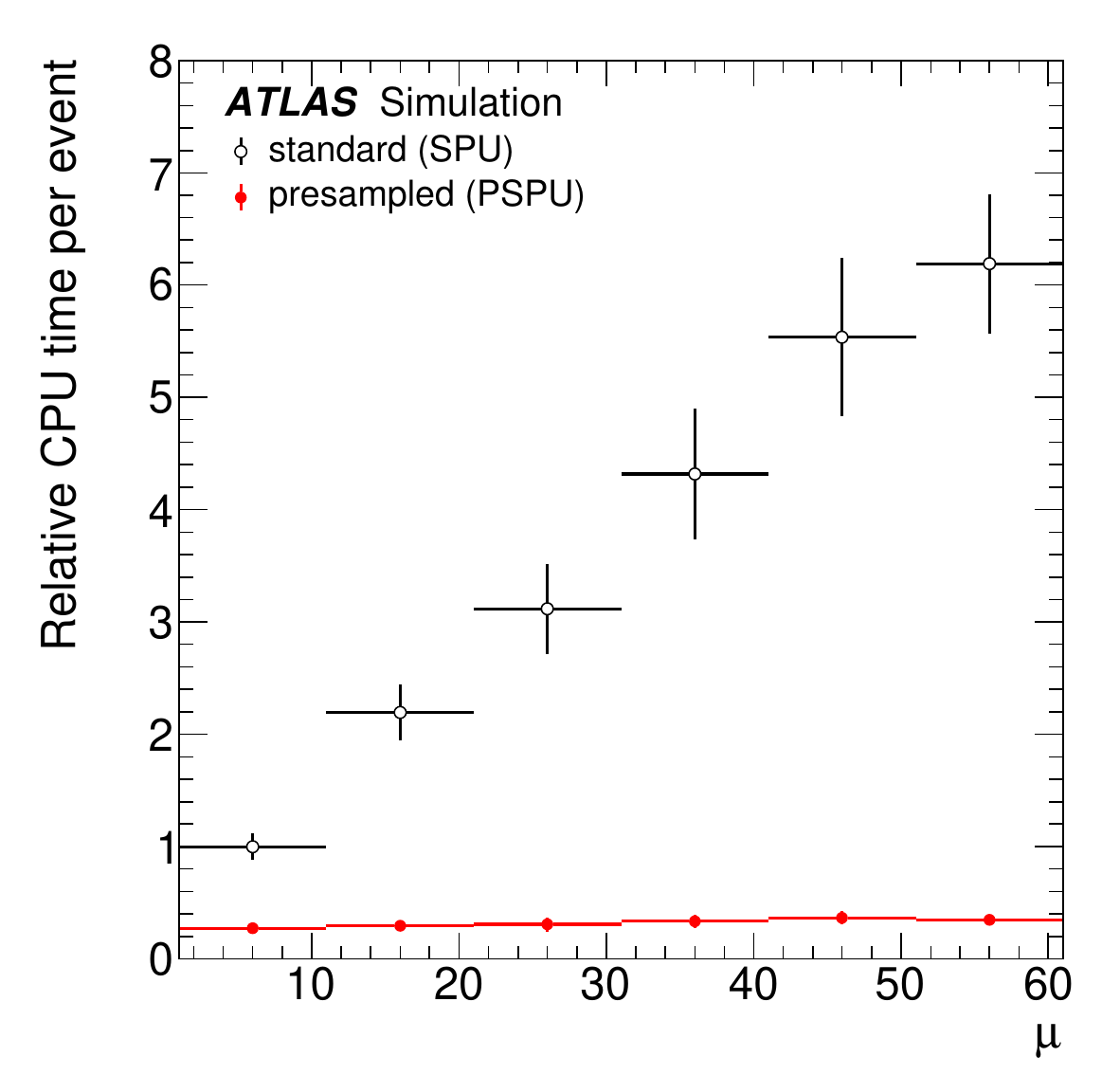}
\caption{Comparison of the average CPU time per event in the \standardmethod (SPU) digitisation
(black open circles) and the \mcoverlay (PSPU) digitisation (red filled circles) as a function of the number of \pp collisions per bunch crossing (\( \mu \)).
The CPU time is normalised to the time taken for the \standardmethod for the lowest \( \mu \) bin.
For this measurement \ttbar events are used for the hard-scatter event.  The average
is taken over 1000 events and the vertical error bars represent the standard deviation of the
separate CPU time measurements. For the \standardmethod digitisation the slope of the relative CPU time per event versus \actualmu
is \(0.108\) while for the \mcoverlay digitisation it is \(0.002\). }
\label{fig:perf_cpu_vs_mu}
\end{figure}

The average CPU time per event in the \standardandover \simcs as a function of
\( \mu \) is shown in \Fig{\ref{fig:perf_cpu_vs_mu}}.  As can be seen, both depend linearly on \( \mu \)
but the slope is about \(50\) times larger
for the \standardmethod than for the \mcoverlay \simc.  For the \standardmethod \simc, the CPU time required at \(\mu=70\) is \(7.5\)
times larger than for \(\mu=10\), while for the \mcoverlay method, the corresponding increase in CPU
time is only a factor of \(1.2\). Extrapolating this to \(\mu=200\), the CPU time is \(20\) times greater
than for \(\mu = 10\) for the standard method and \(<2\) times higher for the \mcoverlay method. However, this comparison does not account for the CPU time required for the production of the presampled pileup dataset, which is needed to assess the overall CPU benefit in a realistic campaign, as is discussed below.
 
\begin{figure}[!htb]
\centering
\includegraphics[width=0.75\textwidth]{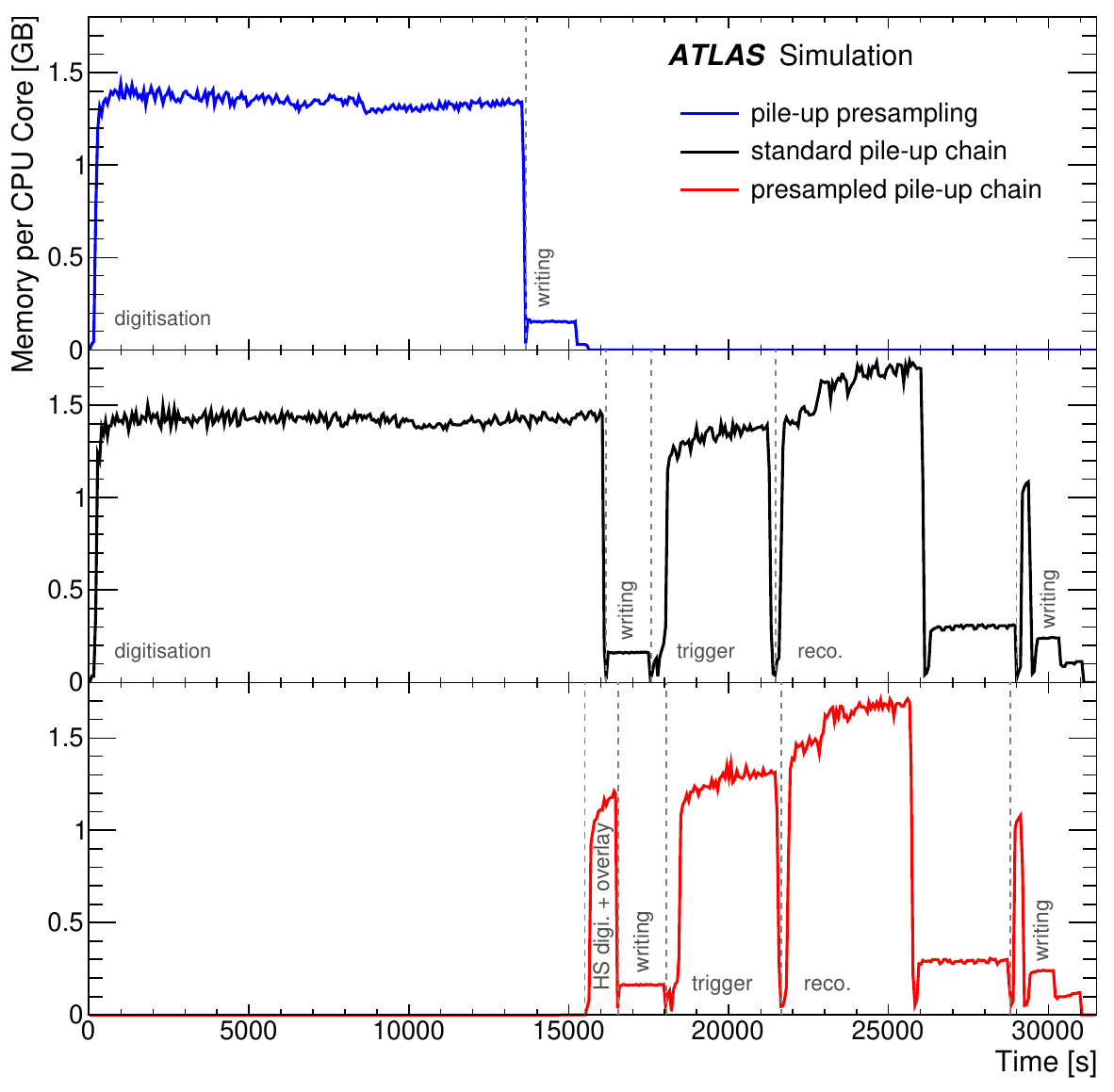}
\caption{The memory usage profile of different production steps as a function of the job wall-time for 2000 hard-scatter events.
The presampling (top), the \standardmethod (middle) and the \mcoverlay (bottom) \simc are compared.
In the latter case, ``HS digi.'' refers to the digitisation of the hard-scatter event.
The underlying \( \mu \) distribution is that corresponding to the 2017 data \( \mu \)
distribution.}
\label{fig:perf_mem}
\end{figure}

\Fig{\ref{fig:perf_mem}} shows the memory used by the various steps as a function of time for the
different production steps for the two simulation chains.
The time estimate is based on running 2000
hard-scatter events for the 2017 \( \mu \) distribution on the same CPU in all cases, so that the three scenarios can be directly compared. The absolute number, of course, depends on the CPU used and the \( \mu \) distribution. The presampling takes about \SI{7}{s} per event. The standard digitisation takes about \SI{8}{s} per event, while the hard-scatter digitisation and overlay of the presampled \pileup takes about \SI{0.5}{s}. The remaining steps, which are the same for the two simulation chains, take about \SI{8}{s} and include the trigger emulation, reconstruction, and the writing of the analysis format to disk.

When comparing the required CPU time between the two chains,
the following equations provide a good approximation. For the \standardmethod \simc, the
time \( T_{\textrm{standard}} \) required is simply given by the number of events in the campaign times the total time
\( t_\textrm{digi} + t_\textrm{other} \), where \( t_\textrm{other} \) is the sum of the times needed for
reconstruction, trigger and writing the event to disk.  Thus
\begin{equation*}
T_\textrm{standard}  = N_\textrm{MC-campaign}\times (t_\textrm{digi} + t_\textrm{other}),
\end{equation*}
where \( N_\textrm{MC-campaign} \) is the number of hard-scatter events produced in a given MC campaign.
 
For the \mcoverlay \simc, the time \( T_\textrm{presample}\) required is given by the number of events in the campaign times the time needed for
the overlay step and other aspects plus the time required for the presampling.  This last contribution is given by the total number of
presampled \pileup events required (\(N_\textrm{pp}\)) multiplied by the event digitisation time, so that the required time is
\begin{equation*}
T_\textrm{presample} = N_\textrm{MC-campaign} \times (t_\textrm{overlay}+t_\textrm{other})+N_\textrm{pp}\times t_\textrm{digi}.
\end{equation*}
 
The time reduction factor of the \mcoverlay \simc compared to the standard is then given by
\begin{eqnarray*}
\frac{T_\textrm{presample}}{T_\textrm{standard}} & = & \frac{N_\textrm{MC-campaign}\times (t_\textrm{overlay}+t_\textrm{other})+N_\textrm{pp}\times t_\textrm{digi} }{N_\textrm{MC-campaign}\times (t_\textrm{other}+t_\textrm{digi}) } \nonumber \\
& \approx & \frac{1}{t_\textrm{other}+t_\textrm{digi}}\left[t_\textrm{other}+t_\textrm{digi} \times  \frac{N_\textrm{pp}}{N_\textrm{MC-campaign}}\right],
\end{eqnarray*}
where the approximation \( t_\textrm{overlay}\ll t_\textrm{other} \) is made, based on the observations from \Fig{\ref{fig:perf_mem}}.
 
It is immediately clear that the \mcoverlay \simc uses less CPU time than the \standardmethod \simc
since \(N_\textrm{pp} < N_\textrm{MC-campaign}\).  Choosing the exact value for \(N_\textrm{pp}\),
however, is not trivial.  In general, the reuse of a given presampled \pileup event within a
particular MC sample, representing an individual hard-scatter physics process, should be avoided if
possible, otherwise each overlaid hard-scatter plus \pileup event would not be statistically
independent. Such \textit{oversampling} would be particularly worrisome if the presampled \pileup
event in question contained a distinctive feature, such as a high-transverse-momentum jet,
which could cause difficulties in using the MC sample for the statistical interpretation of the data distributions.
In practice, such a repetition would not be statistically significant in the bulk of a distribution
but could be problematic in the tails, where there are few events.  Given this, it is reasonable
that the value for \(N_\textrm{pp}\) be chosen to be about the size of the largest individual MC
sample, so that no event is repeated within it.
 
For the ATLAS \RunTwo MC campaign, \(N_\textrm{MC-campaign}\sim \SI{10}{B}\) and the single largest individual
MC sample had a size of \SI{0.2}{B} events.  Allowing for some increase in these sizes to be commensurate
with the size of the evolving data samples, \( N_\textrm{pp} \sim \SI{0.5}{B} \) should thus be sufficient.
Taking the resulting \( N_\textrm{MC-campaign} / N_\textrm{pp} \sim 20 \), along with \( t_\textrm{other}\approx
t_\textrm{digi}\) (as seen in \Fig{\ref{fig:perf_mem}}), the ratio of the times required for the two
methods is \(T_\textrm{presample}/T_\textrm{standard}\sim 0.53\).  Hence, the \mcoverlay \simc
provides a CPU saving of 47\% for the combined digitisation, reconstruction, trigger and writing steps, compared to the \standardmethod \simc.  If the time required for
reconstruction and trigger is further improved (as is planned for \RunThr), or the digitisation time were to further increase due to \pileup, the ratio would decrease; e.g.\ if
\( t_\textrm{other}\approx t_\textrm{digi}/2 \), a CPU saving of 63\% would be realised.
The \( \sim 50\% \) reduction translates to an overall MC production CPU saving of around 20\%, since 60\% of the CPU time required for the \simc is at present used for event generation and G4 simulation, which is not affected by this improvement.
These are illustrative examples that confirm the intuitive expectation that performing the digitisation
just once per campaign is much more effective than doing it for each simulated hard-scatter event,
as the number of presampled events needed is by construction smaller than the number of
hard-scatter events.

From the memory usage point of view, the \mcoverlay load is similar to the \standardmethod and well below the (soft) production limit of \(\sim \SI{2}{GB} \) per core (see \Fig{\ref{fig:perf_mem}}) for the \( \mu \) values observed during \RunTwo and expected for \RunThr.
However, compared to the \standardmethod, the \mcoverlay \simc puts less
stress on the I/O system both because, as is mentioned above, the presampled \pileup dataset files are
about a factor of five smaller and because they can be read sequentially. The sequential reading is
possible because the random access necessary to combine the minimum-bias input files in
the \standardmethod is now performed only once at the presampling stage.  Hence, the
presampled \pileup RDO production, with its heavier requirements, can be performed on a limited
subset of ATLAS MC production sites designed to cope well with such workloads; the subsequent \mcoverlay \simc will then run on all resources available to ATLAS, utilising the \( \approx 20\% \) of sites that have previously been excluded for reconstruction due to insufficient I/O or disk resources.  The reduced input size also enables the usage of opportunistic resources such as high-performance computing (HPC) sites, which typically have less available storage for ATLAS.\@  The smaller I/O requirements from
the \mcoverlay \simc jobs simplify the production workflow, and make it possible to transfer
the \pileup datasets on demand to the computing node at a given production site, where they are
needed. If network speed is further increased in the future, it might even be possible to access
them directly via the network during the job from a remote storage site.
 
The Analysis Object Data (AOD) event size written to disk is the same for both methods, i.e.\ there is neither advantage
nor disadvantage in using the \mcoverlay \simc in this regard. However, the many simulated minimum-bias
events do not have to be distributed as widely any more throughout the year as they only need to be
accessed once for creating the presampled events. These presampled events need to be made available
widely though.  It is expected that these two effects roughly cancel out but operational experience
is needed to understand how to distribute the presampled sample in the most effective way.
 
\section{Inner detector}
\label{sec:indet}
The ID consists of three subdetectors which all use different technologies as discussed in \Sect{\ref{sec:det}}. Each of them has separate digitisation software and hence a different treatment for the \mcoverlay procedure is required for each. In this section, the readout of the three ID subdetectors is described, along with the \mcoverlay procedure
for each. Validation results are also presented.
 
\subsection{Detector readout}
\paragraph{Silicon Pixel detector:}
The charge produced by a particle traversing a silicon pixel is integrated if it passes a set
threshold.  In \RunTwo, this threshold is typically around 2500 electrons for the IBL and 3500
electrons for the remainder of the Pixel detector.  The resulting charge deposited by a minimum-ionising
particle (MIP) that traverses a single pixel is typically 16\,000 and 20\,000 electrons,
respectively.  The amount of charge deposited by a particle traversing the detector varies depending
on the path length of the particle through the active silicon and can be spread across multiple
pixels.  The length of time during which the charge signal exceeds the threshold, termed
\textit{time-over-threshold} (ToT), is recorded. The ToT is roughly proportional to the charge. While most
of the charge drifts to the pixel readout within the \SI{25}{ns} bunch crossing time of the LHC,
there is a small fraction which may take longer and only arrive in the subsequent bunch crossing
(BC+1). Thus, in any given bunch crossing, the \pileup events both from the previous and the current
bunch crossings contribute hits.
 
\paragraph{Silicon microstrip detector (SCT):}
For the SCT, the readout is in principle similar to the Pixel detector in that a threshold is
applied for each strip. But, in contrast to the pixel readout, it is purely digital, i.e.\ neither
the charge nor the ToT is stored for a given strip, just a bit, X = 0 or 1, to signal a hit (1) or
the absence of a hit (0). Hence, the hit from the current BC as well as that of the two adjacent
bunch crossings (i.e.\ BC--1 and BC+1) are read out. Several data compression modes have been used
since the first LHC collisions; they are defined by the hit pattern of the three time bins:
\begin{itemize}
\item{Any-hit mode (1XX, X1X or XX1); channels with a signal above threshold in either the current, previous or next
bunch crossing are read out. }
\item{Level mode (X1X); only channels with a signal above threshold in the current bunch crossing are
read out.}
\item{Edge mode (01X); only channels with a signal above threshold in the current bunch crossing and
explicitly no hit in the preceding bunch crossing are read out.}
\end{itemize}
The data can be compressed further by storing, for adjacent strips with hits above threshold, only the address of the first
strip and the number of these adjacent strips. When this compression is invoked, the information about which of the three bunch crossings observed a hit for a given strip is lost.
When the LHC is
running with \SI{25}{ns} bunch spacing, SCT RDOs are required to satisfy the 01X hit pattern to be
considered during event reconstruction in order to suppress \pileup from the previous crossings.
 
\paragraph{Transition radiation tracker (TRT):}
When a particle crosses one of the tubes in the TRT, the electrons drift to the anode wire, producing an electrical signal. If the charge of that signal exceeds a low discriminator threshold, a corresponding hit is recorded, in eight time slices of \SI{3.125}{ns} each.  The drift time is calculated based on the time of the first hit, which is subsequently converted to  distance to give a drift-circle radius.  In addition, in order to provide information for electron identification, a record is kept of whether a high discriminator threshold is exceeded in any of the eight time slices. This information is stored for the previous, current and subsequent bunch crossings (i.e.\ BC--1, BC, BC+1).

\subsection{Overlay procedure}
\label{sec:overlayproc}
The quantities which are overlaid for the inner detector are the RDOs. Due to the high number of channels in the inner detector, zero
suppression\footnote{Rather than reading out all channels, only those channels containing data
(above a certain significance level) are recorded.} is employed to reduce the amount of data read out
and stored from the detector.
Since for the ID the RDOs do not contain the full information of the HITS created by simulation, the overlay of
RDO information is less accurate than the overlay of the underlying HITS information. However, the
impact on physics observables is generally found to be negligible as is described in the following;
where a difference is observed, a parameterised correction is derived as is described below.

\paragraph{Pixel detector:}
The Pixel detector has in excess of \SI{90}{M} readout channels and a very high granularity. The
single-pixel occupancy is below \(2.5\times 10^{-5}\) per unit \actualmu in all layers~\cite{pixeliblocc},
so even at \(\actualmu \sim 100\) it is below 0.25\%.  Therefore, the chance that a single pixel which
contains a signal due to a charged particle from the hard-scatter event also contains one from the
overlapping in-time \pileup events is \( < 0.25\% \). A pixel RDO contains the channel identifier and a
32-bit packed word containing the ToT, a bunch-crossing identifier, and information related to the
L1 trigger not relevant in simulation.  In the \mcoverlay, if an RDO of a given channel
contains a hit above threshold from either the hard-scatter event or the \pileup event, but not both, the
corresponding RDO is kept and written out. In the \( \lesssim 0.25\% \) of cases where it contains a hit
above threshold in both the hard-scatter event and the \pileup event, only the hard-scatter RDO is kept in
order to retain the ToT (and thus, for example, the energy deposited per path length \(\dif E / \dif x\)) from
the signal process. This causes a small loss of information as in principle the ToT would be
modified by the presence of the additional charge deposited in that pixel from the \pileup events.
But, as it only affects a small fraction of cases, it has a negligible impact on the overall
physics performance.  In addition, there could be a loss of information if, for a given pixel, both
the hard-scatter event and the \pileup event produce charge deposits which are below the readout threshold
but whose sum is above the threshold.  In this case the \mcoverlay method will register no hit
while the standard method will register a hit above threshold. This effect could
reduce the cluster size and the ToT.  But again, only a very small fraction of pixels
are affected, so both the cluster size and the ToT agree well between the two methods.
 
\paragraph{SCT detector:}
The SCT is a strip detector with \SI{6.3}{M} readout channels and an occupancy in high \pileup conditions
of \( \mathcal{O}(1\%) \); consequently the \pileup modelling is more critical than for the pixel
detector. In order to facilitate accurate modelling, it is important that presampled RDOs be
stored in any-hit mode, without further compression, to ensure that the impact of
out-of-time \pileup is modelled correctly. To combine hard-scatter and \pileup RDOs, all of the
strips that are hit on a module are unpacked from the respective RDOs and repacked into RDOs using
the desired compression mode. Loss of information only occurs if hits in both the hard-scatter event and the
\pileup event are below threshold but the sum of the two charges is above threshold. In this
case, in the standard digitisation a hit would be present while with the \mcoverlay procedure it is
not, causing the \mcoverlay procedure potentially to result in fewer SCT hits per track. The impact
is, however, negligible as is shown below.
 
\paragraph{TRT detector:}
The TRT is a straw tube detector with \SI{320}{k} readout channels, and in high \pileup conditions the
occupancy of the TRT exceeds 10\%. Therefore, \pileup has a major impact on the TRT
signals. If the channel identifiers in the hard-scatter and \pileup events are the same, the data word
stored is set to a bit-wise logical OR of the corresponding raw words. This
results in some loss of information as the sum of the charge signals will be larger, and thus more
easily pass a given threshold, than would be just the sum of the digitised signals. This particularly impacts
the fraction of hits that pass the high discriminator threshold.
 
A correction for this effect is applied to improve the level of agreement between the \mcoverlay and the
standard digitisation. For this correction, a high-threshold (HT) bit is activated according to a
randomised procedure, tuned to describe the standard digitisation. The rate of randomly activating a
high-threshold bit is parameterised as a linear function of the occupancy of the TRT in the
simulated \pileup events (a proxy for the average energy deposited in the \pileup events) and
whether the charged particle that is traversing the straw from the hard-scatter event is an electron
or not.  A different correction is applied for electrons as they produce significant amounts of
transition radiation in the momentum range relevant for physics analysis (5--\SI{140}{\GeV}), while
all other particles do not.  The correction corresponds to approximately a 10\% (5\%) increase in
the number of HT hits for electrons (non-electrons) at the average \RunTwo \actualmu value.

\subsection{Validation results}
 
To validate the \mcoverlay digitisation for each of the subdetectors, the properties of tracks in
simulated \ttbar events, where at least one \(W\) boson from the top quarks decays leptonically,
are compared between the \mcoverlay method and the standard digitisation. The \ttbar events
are chosen because they represent a busy detector environment and contain tracks from a wide range of
physics objects.
 
The primary track reconstruction is performed using an iterative track-finding procedure seeded from
combinations of silicon detector measurements.  The track candidates must have a transverse momentum
\( \pT > \SI{500}{\MeV} \) and \(\abseta < 2.5\) and meet the following criteria: a minimum of seven pixel
and SCT clusters, a maximum of either one pixel or two SCT clusters shared among more than one
track and no more than two holes\footnote{A hole is defined as the absence of a hit on a traversed
sensitive detector element.} in the SCT and Pixel detectors combined.  The tracks formed from the silicon detector measurements are then
extended into the TRT detector.  Full details, including a description of the TRT track extensions,
can be found in \Refns{\cite{newt,PERF-2015-08}}. 
 
\begin{figure}[htb]
\centering
\subfloat[]{\includegraphics[width=0.45\textwidth]{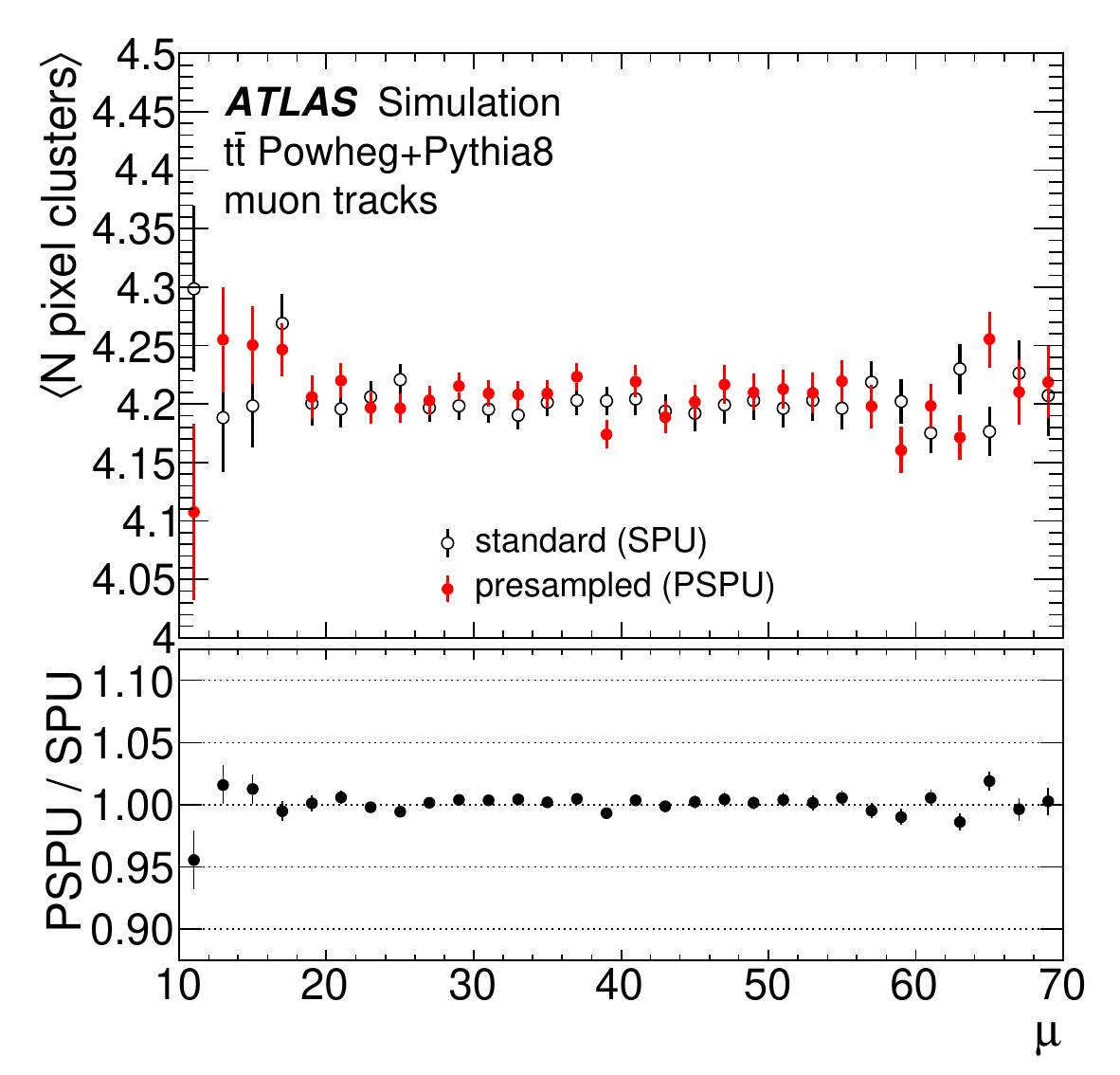}}
\subfloat[]{\includegraphics[width=0.45\textwidth]{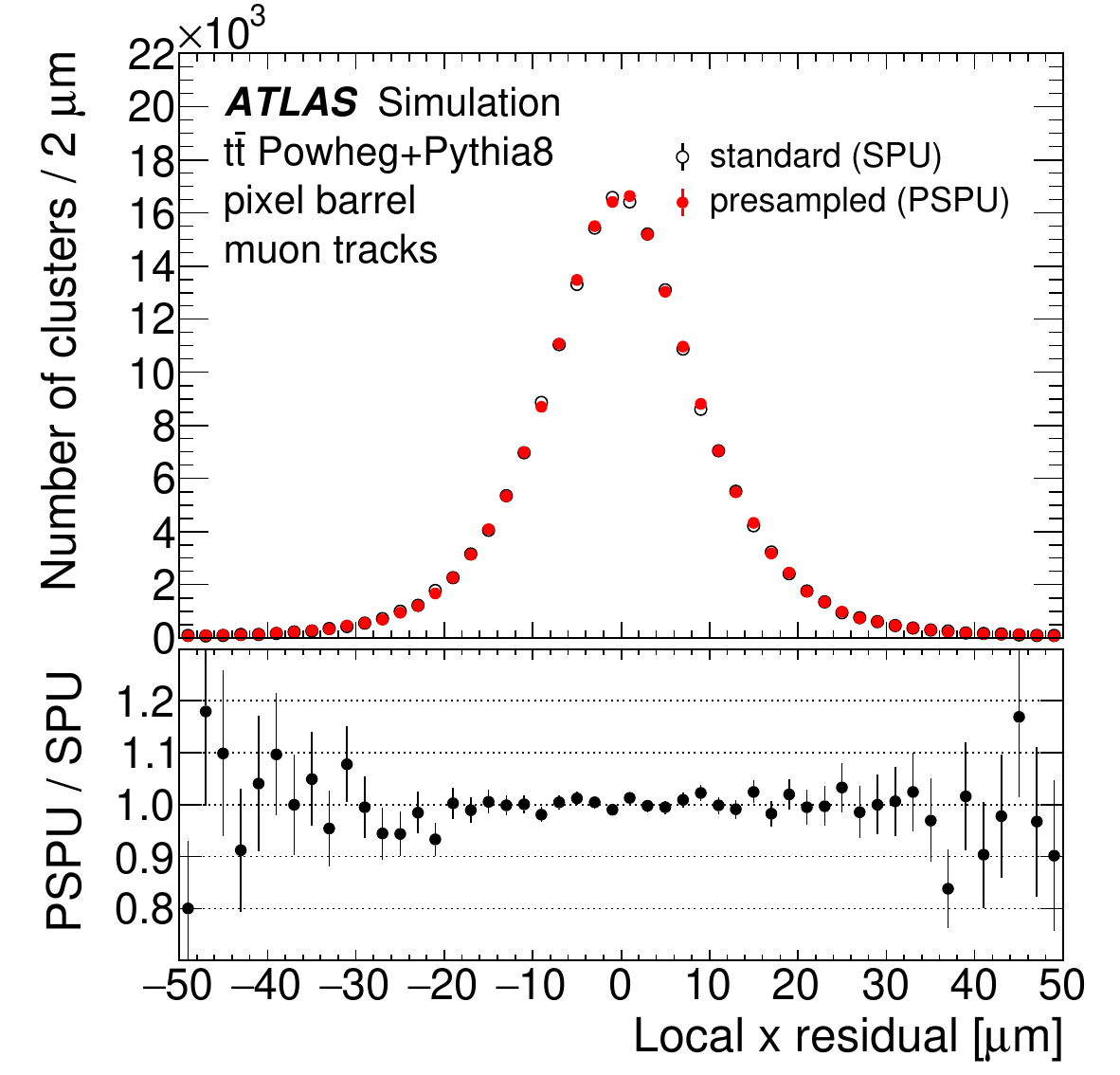}}
\caption{Comparison of between the standard digitisation (open black circles) and the \mcoverlay (red filled circles), showing (a) the average number of pixel clusters on a track as a function of \actualmu and (b) the the local \(x\) residuals, for tracks produced by muons in simulated \ttbar events. The distributions are integrated over all clusters associated with muon tracks in the hard-scatter event. The residual is defined as the measured hit position minus the expected hit position from the track extrapolation (not including the cluster in question). The bottom panels show the ratios of the two distributions.}
\label{fig:ID_Muon_Pix}
\end{figure}

\begin{figure}[htb]
\centering
\subfloat[]{\includegraphics[width=0.45\textwidth]{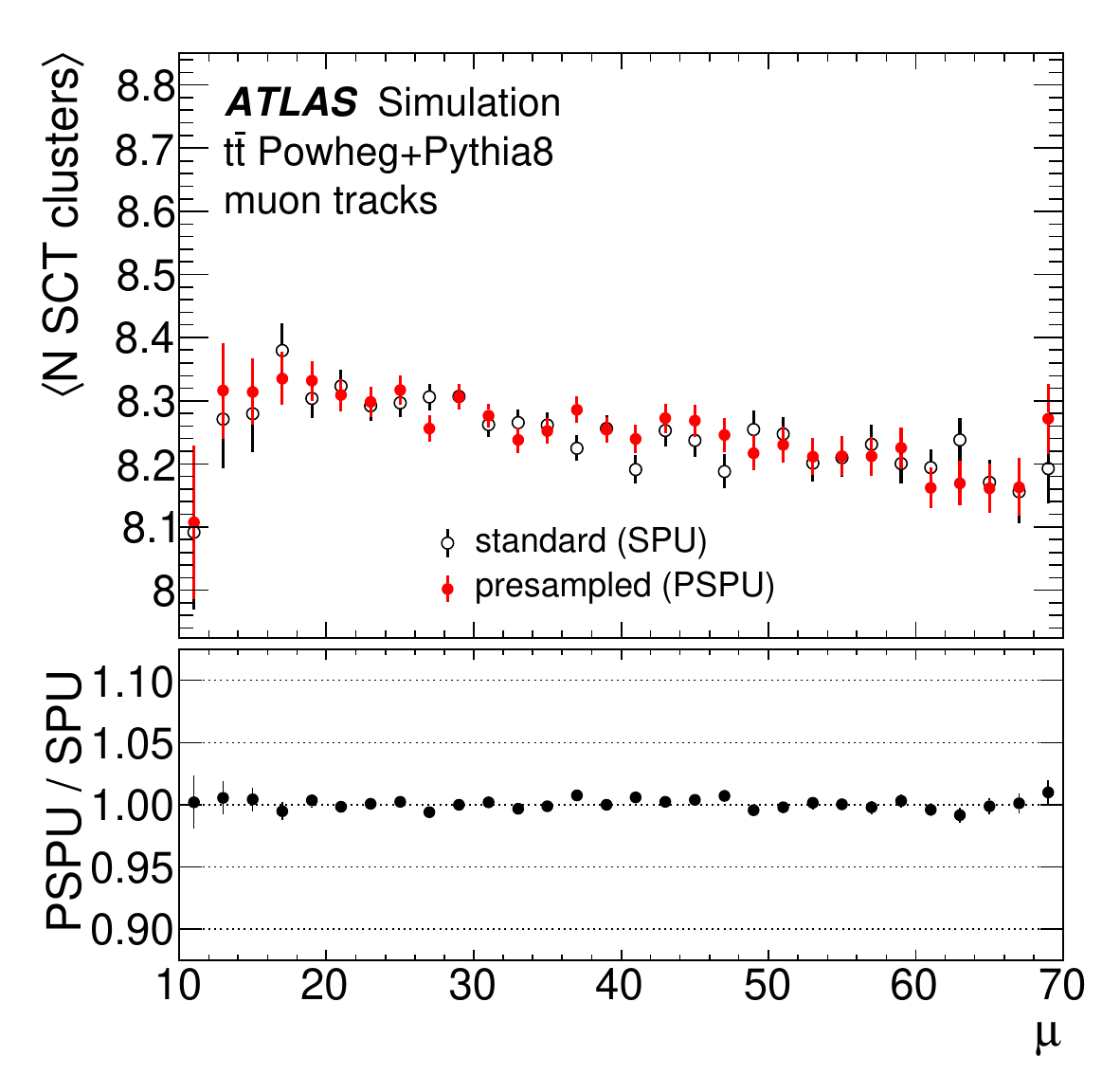}}
\subfloat[]{\includegraphics[width=0.45\textwidth]{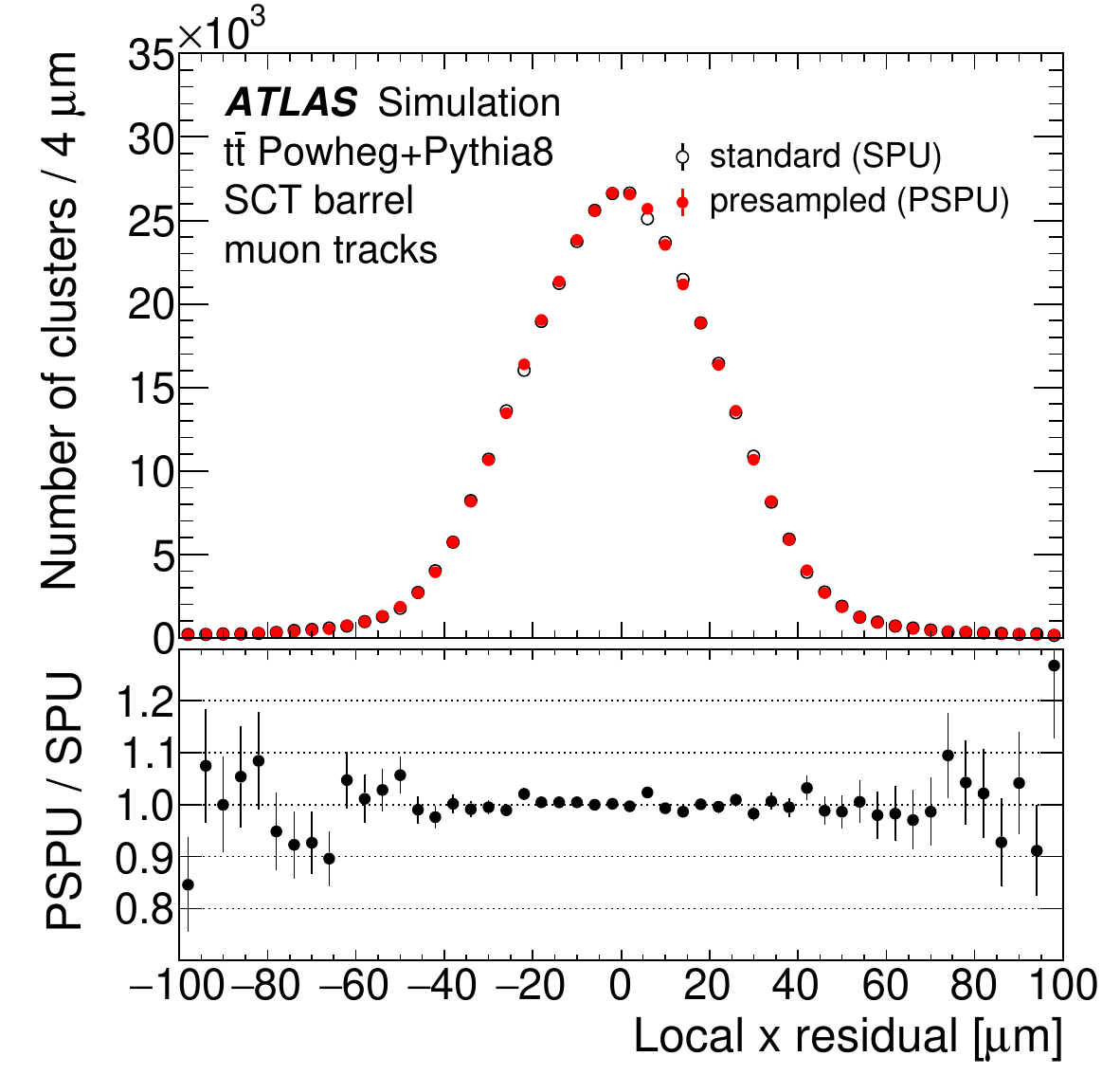}}
\caption{Comparison of between the standard digitisation (open black circles) and the \mcoverlay (red filled circles), showing (a) the average number of SCT clusters on a track as a function of \actualmu and (b) the the local \(x\) residuals, for tracks produced by muons in simulated \ttbar events. The distributions are integrated over all clusters associated with muon tracks in the hard-scatter event. The residual is defined as the measured hit position minus the expected hit position from the track extrapolation (not including the cluster in question). The bottom panels show the ratios of the two distributions.}
\label{fig:ID_Muon_SCT}
\end{figure}

\Fig{\ref{fig:ID_Muon_Pix}} shows the number of pixel clusters associated with a muon track as a function
of \actualmu, and the unbiased residual in the local \(x\) coordinate, which corresponds to the direction with the
highest measurement precision.  The unbiased residual is the distance of the cluster from the track
trajectory (not including the cluster itself) at the point where that trajectory crosses the pixel sensor.
\Fig{\ref{fig:ID_Muon_SCT}} shows the corresponding quantities for the SCT.\@ In all cases,
the \mcoverlay and standard digitisation are shown, and good agreement is observed between the two
methods.
 
\begin{figure}[htb]
\centering
\subfloat[]{\includegraphics[width=0.45\textwidth]{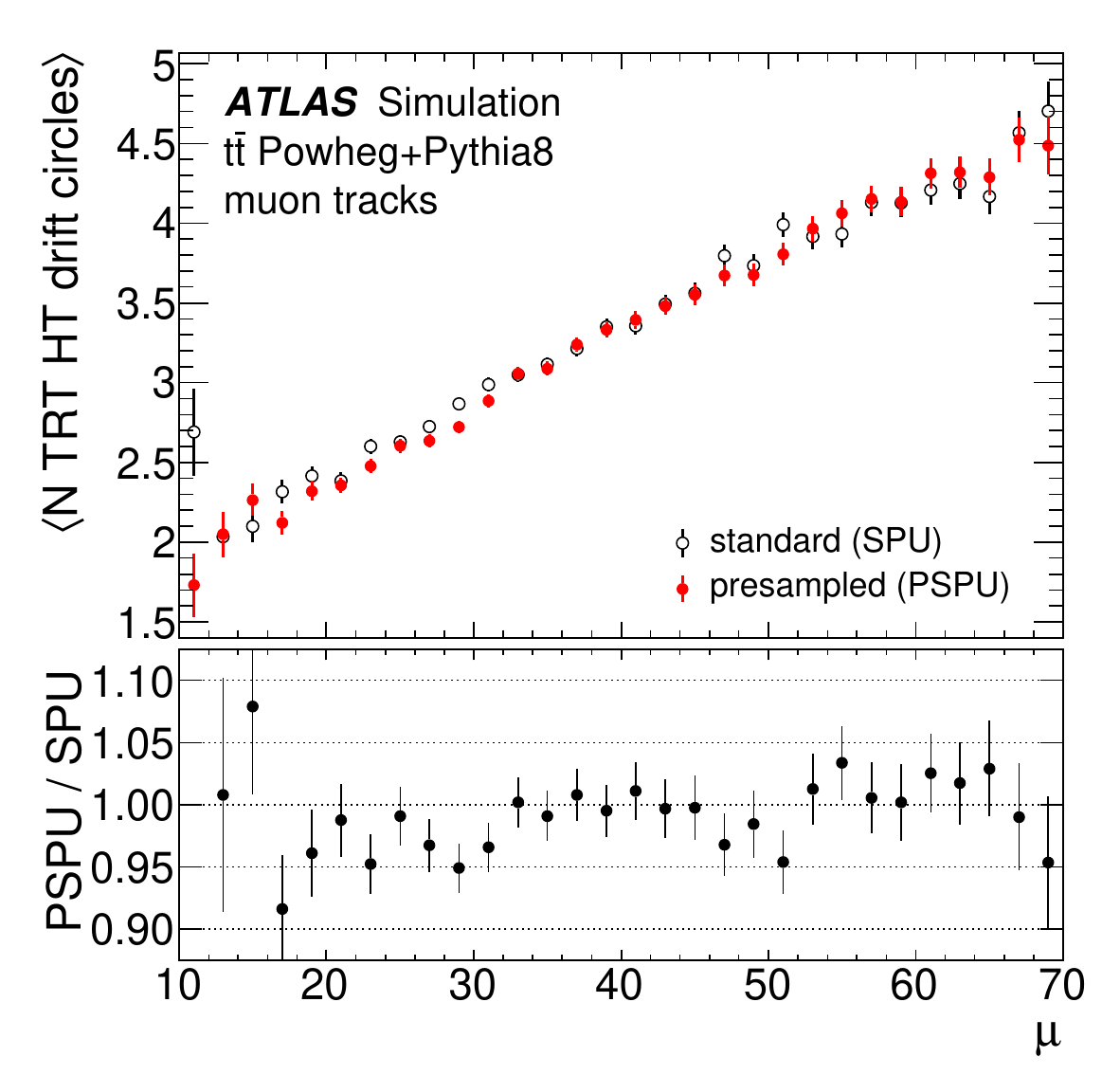}}
\subfloat[]{\includegraphics[width=0.45\textwidth]{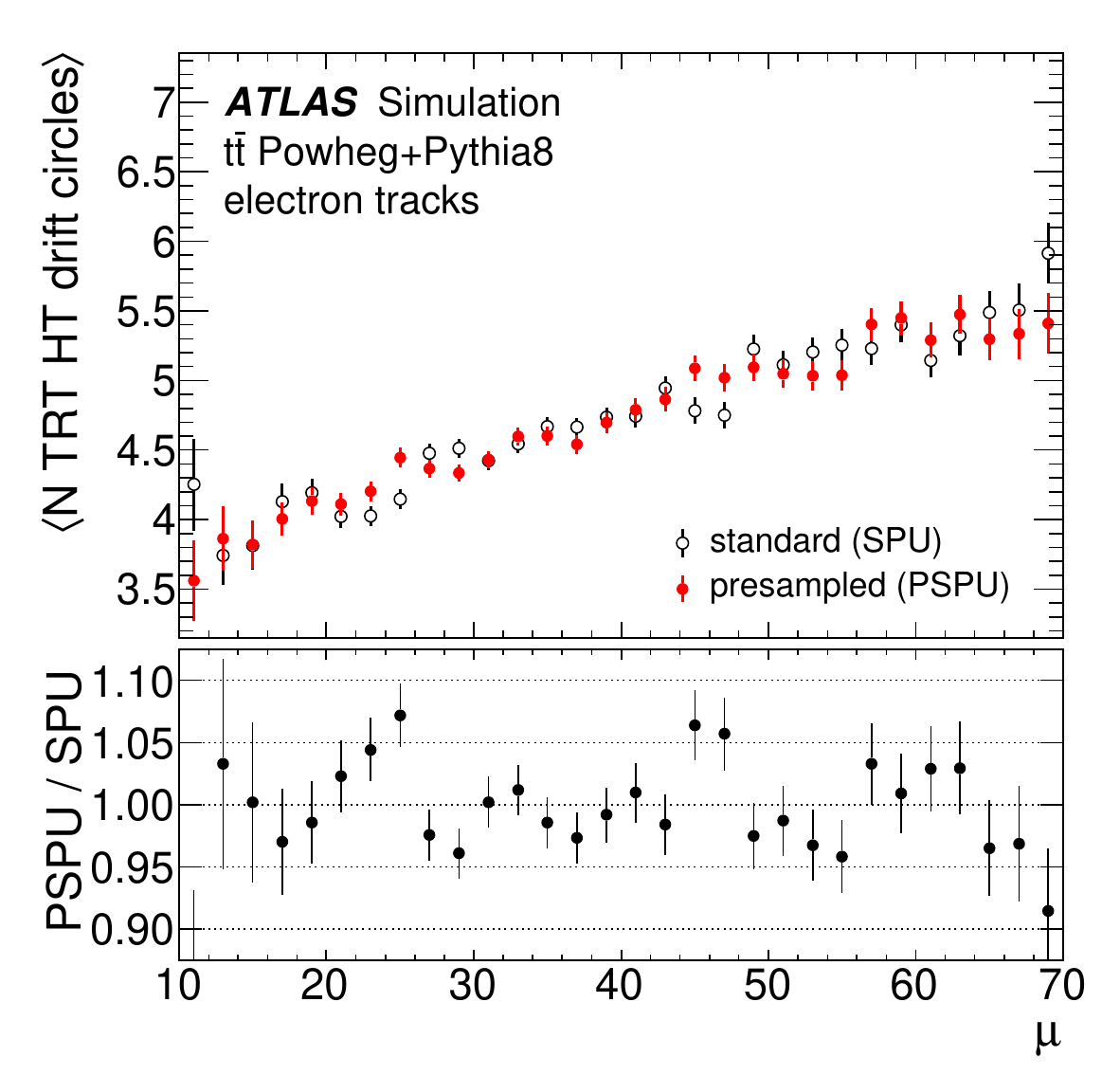}}
\caption{Distributions of the average number of TRT high-threshold drift circles, after the corrections described in the text, for tracks produced by (a) muons and (b) electrons in simulated \ttbar events as a function of \actualmu. The standard digitisation (open black circles)
is compared with the \mcoverlay (red filled circles). The bottom panels show the ratios of the two distributions.}
\label{fig:ID_TRT_HT}
\end{figure}

\Fig{\ref{fig:ID_TRT_HT}} shows a comparison of the number of high-threshold TRT drift circles as a
function of \actualmu for muons\footnote{ The proportion of transition radiation, and hence
high-threshold hits for pions, which dominate in \pileup events, will behave similarly to muons due
to their comparable mass.  Muons are chosen here because their momentum spectrum in \ttbar events
is comparable to that of the electrons and hence allow a direct comparison.} and electrons. As is
explained above, due to the high occupancy of the detector, the number of high-threshold drift
circles is particularly sensitive to the \mcoverlay procedure.  After the parameterised corrections
discussed in \Sect{\ref{sec:overlayproc}} are applied, the average numbers of high-threshold drift
circles for electrons and muons are each comparable for the two methods.
 
\begin{figure}[htb]
\centering
\subfloat[]{\includegraphics[width=0.45\textwidth]{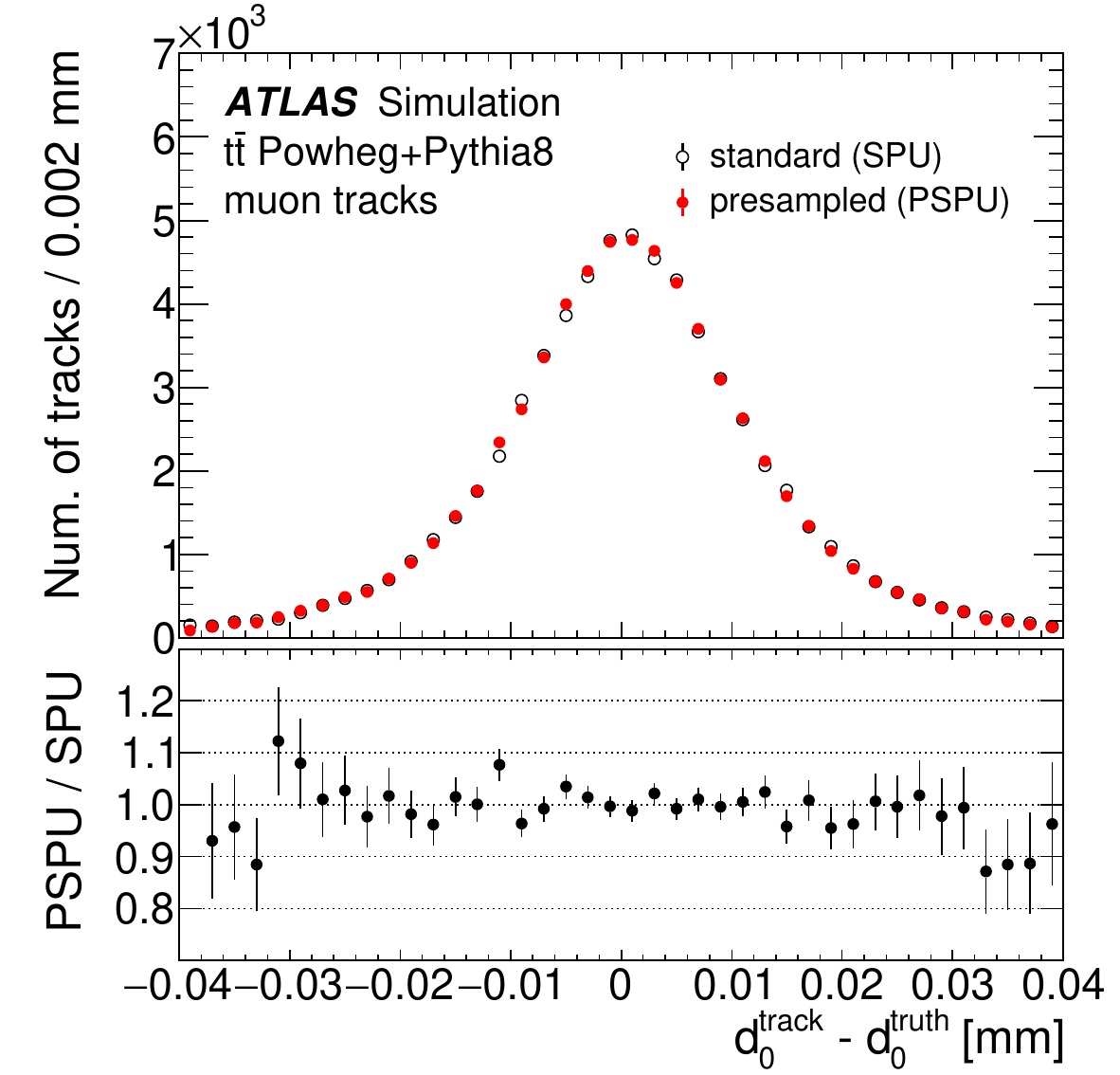}}
\subfloat[]{\includegraphics[width=0.45\textwidth]{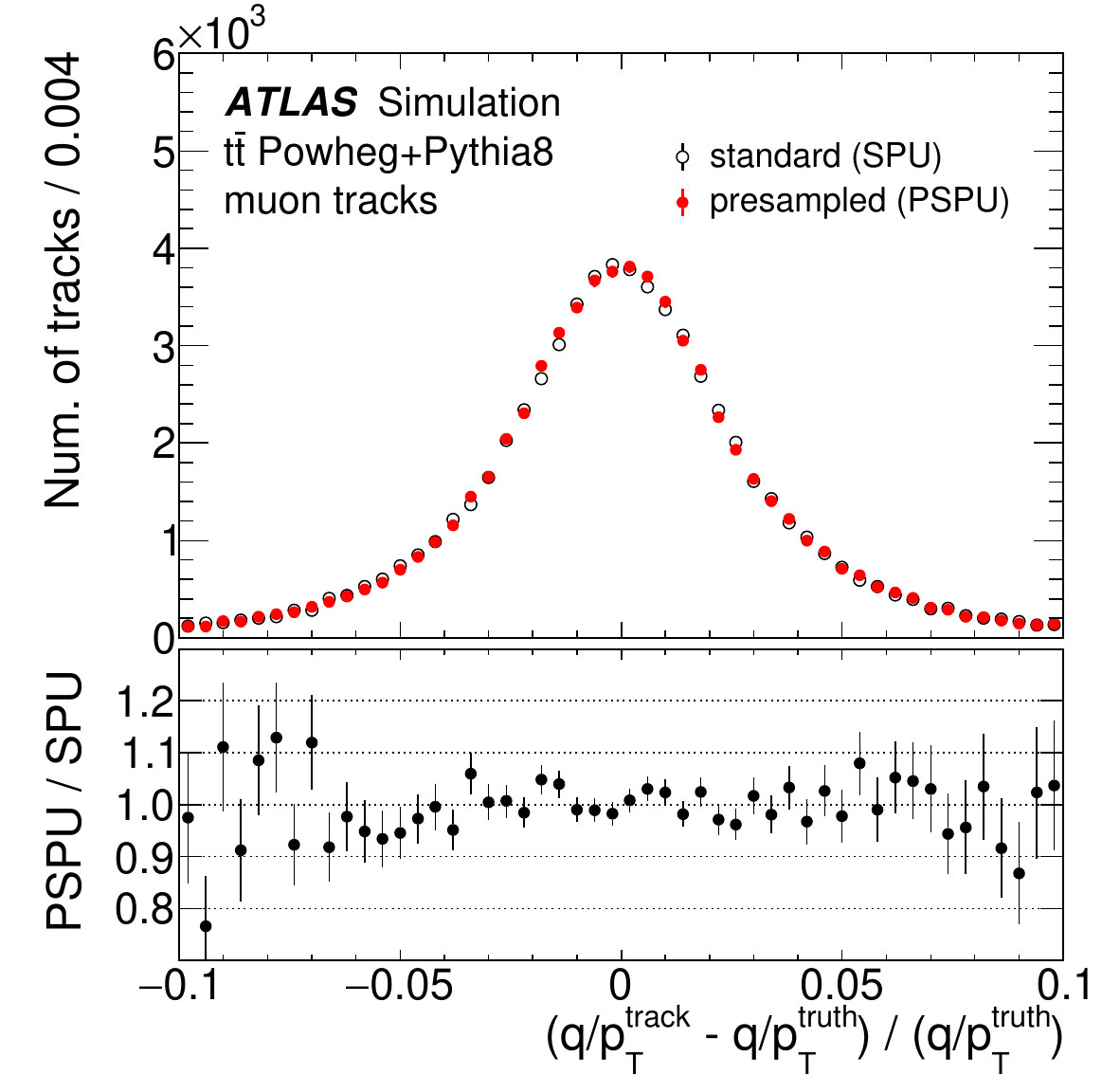}}
\caption{A comparison of the reconstructed muon track parameter resolution for  (a)
\dzero and (b) \( q/\pt \) between the standard digitisation (open black circles) and the \mcoverlay
(red filled circles) methods, for simulated \ttbar events. The bottom panels show the ratios of the two distributions.}
\label{fig:ID_Reso_ttbarA}
\end{figure}

\begin{figure}[htb]
\centering
\subfloat[]{\includegraphics[width=0.45\textwidth]{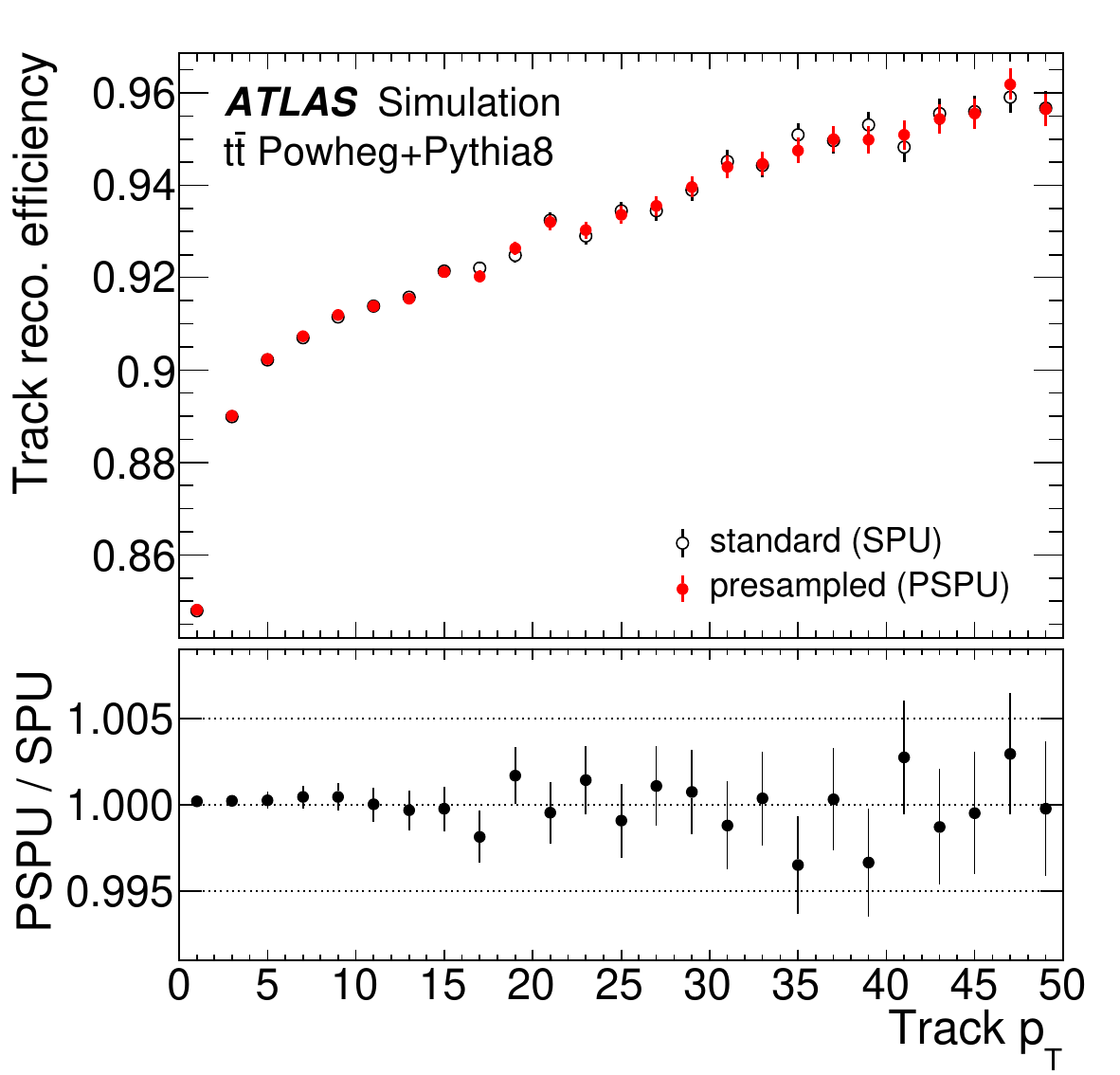}}
\subfloat[]{\includegraphics[width=0.45\textwidth]{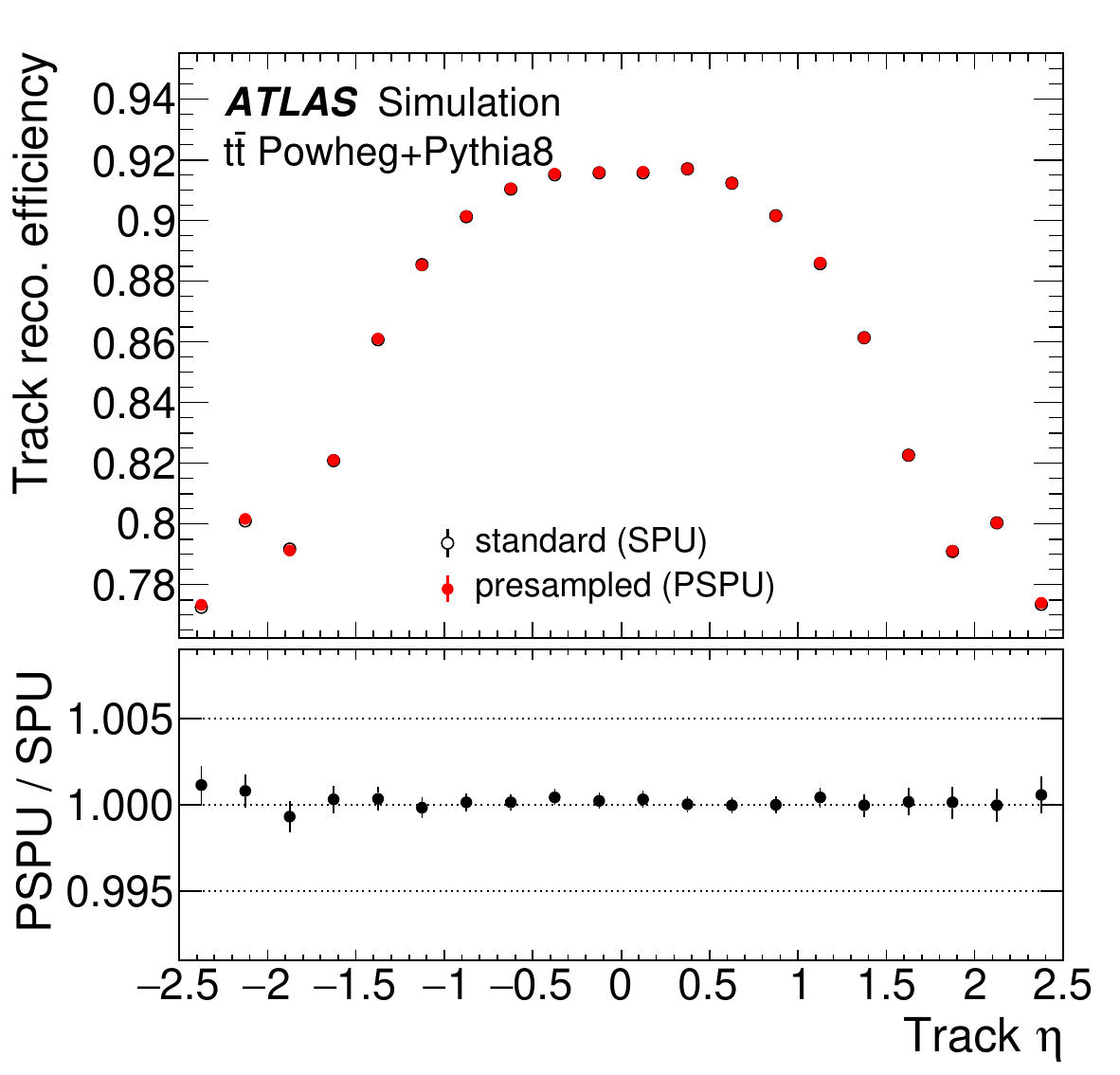}}
\caption{A comparison of the track  reconstruction efficiency between the standard digitisation (open black circles) and the \mcoverlay
(red filled circles) methods, for simulated \ttbar events as a function of (a) transverse momentum and (b)
pseudorapidity (\( \eta \)). All primary charged particles with \(\pt > \SI{500}{\MeV}\) and \(\abseta < 2.5\) are selected. The bottom panels show the ratios of the two distributions.}
\label{fig:ID_EFf_ttbar}
\end{figure}

The resolution of all track parameters was examined for both methods, and they were found to agree well. \Fig{\ref{fig:ID_Reso_ttbarA}} shows the difference between the reconstructed and true
values for the impact parameter of the track relative to the primary vertex (\dzero), measured in the transverse plane, and the track curvature
(\(q/\pt^\textrm{track}\)) for muons in \ttbar events.
Finally, the track reconstruction efficiency is shown in \Fig{\ref{fig:ID_EFf_ttbar}} as a function
of the \pt and \( \eta \) of all tracks identified in \ttbar events.  The level of agreement between the two
methods is better than 0.5\%.
 
\clearpage

 
\section{Calorimeters}
\subsection{Detector readout}
The \standardandover digitisation algorithms are based on an accurate emulation of the readout of the calorimeter system.
 
For the LAr calorimeter~\cite{LARG-2009-01}, the deposit of energy in the liquid-argon gaps induces
an electric current proportional to the deposited energy. For a uniform energy deposit in the gap,
the signal has a triangular shape as a function of time with a length corresponding to the maximum
drift time of the ionisation electrons, typically \SI{450}{ns} in the EM calorimeter.  This signal
is amplified and shaped by a bipolar \( CR\)--\( {(RC)}^2 \) filter in the front-end readout
boards~\cite{1748-0221-5-09-P09003} to reduce the effect of out-of-time \pileup energy deposits from
collisions in the next or previous bunch crossings.  To accommodate the required dynamic range,
three different gains (high, medium and low) are used. The shaped and amplified signals are sampled
at the LHC bunch-crossing frequency of \SI{40}{MHz} and, for each L1 trigger, are digitised by a
12-bit analog-to-digital converter (ADC). The medium gain for the time sample corresponding to the
maximum expected amplitude is digitised first to choose the most suitable gain for a given
signal. Four time samples for the selected gain are then digitised and sent to the back-end
electronics via optical fibres. For the EMB, EMEC and FCAL calorimeters, the position of the maximum
of the signal is in the third time sample for an energy deposit produced in the same bunch crossing as
the triggered event. For the HEC, it is in the second time sample.
 
For the Tile calorimeter~\cite{TCAL-2017-01}, each cell is read out by two photomultiplier channels.
The maximum height of the analogue pulse in a channel is proportional to the amount of energy
deposited by the incident particle in the corresponding cell.  The shaped signals are sampled and
digitised by 10-bit ADCs at a frequency of \SI{40}{MHz}.  The sampled data are temporarily stored in a
pipeline memory until an L1 trigger signal is received.  Seven time samples, centred around the pulse
peak, are obtained.  A gain selector is used to determine which gain information is sent to the
back-end electronics for event processing.  By default the high-gain signal is used, unless any of
the seven time samples saturates the ADC, at which point the low-gain signal is transmitted.
 
\subsection{Overlay procedure}
 
The procedure for the LAr calorimeter is described in detail below; a very similar procedure is used for the Tile calorimeter.
 
In the presampled RDO sample, the pulse shape (ADC data vs time sample) is stored over the time period for which the calorimeter is read out for each calorimeter cell without any zero suppression.
Its computation is based on the standard \pileup simulation,
described in more detail in \Refn{\cite{ATL-LARG-PUB-2007-011}}.  It considers the energy deposited in each cell for each bunch crossing over the time window affecting the triggered BC, taking into account the time of each event relative to the trigger time.
The resulting pulse shape, expressed in energy versus time, is then converted to ADC counts, applying the energy-to-ADC calibration factor per cell and adding the ADC pedestal.
The gain used in the readout electronics
for this conversion is selected by emulating the logic applied in the front-end readout electronics.
The electronics noise is then added to the presampled RDO, with the proper correlation of the noise between the different
samples, with a value that depends on the gain used to digitise the pulse.
 
In the \mcoverlay step, the pulse shape of the presampled event is converted back into energy and
then the energy from the hard-scatter event is added. This is done for each time sample, resulting
in a combined pulse shape of the hard-scatter and presampled \pileup events.  From this summed pulse shape, the
energies in each time sample are then converted back to ADC counts to produce a pulse shape
mimicking the output of the front-end electronics. The readout electronics gain used in this
conversion is selected according to the energies of the summed pulse shape.  If this gain differs from
the ones used in the hard-scatter or presampled samples, the electronics noise is corrected accordingly.

This pulse shape is then processed following exactly the same algorithm as used in the standard \pileup
digitisation, applying the optimal filtering coefficients~\cite{Cleland:2002rya} to estimate the
energy per cell~\cite{ATL-LARG-PUB-2007-011}.  For cells with high enough energy, the time and pulse
quality factors are also computed.
 
Since all cells are stored in the presampled RDO sample without any suppression, and the energy
response is perfectly linear in the digitisation, the \mcoverlay does not rely on any approximations
except for the integer rounding that is applied when storing ADC counts in the presampled sample. In
practice, the impact of ADC integer rounding was found to be almost negligible.  This rounding
effect only applies to the LAr case; Tile ADC data are actually stored as floats in the
presampled RDO sample.
 
\subsection{Validation results}
 
\begin{figure}[htb!]
\centering
\subfloat[]{\includegraphics[width=0.45\textwidth]{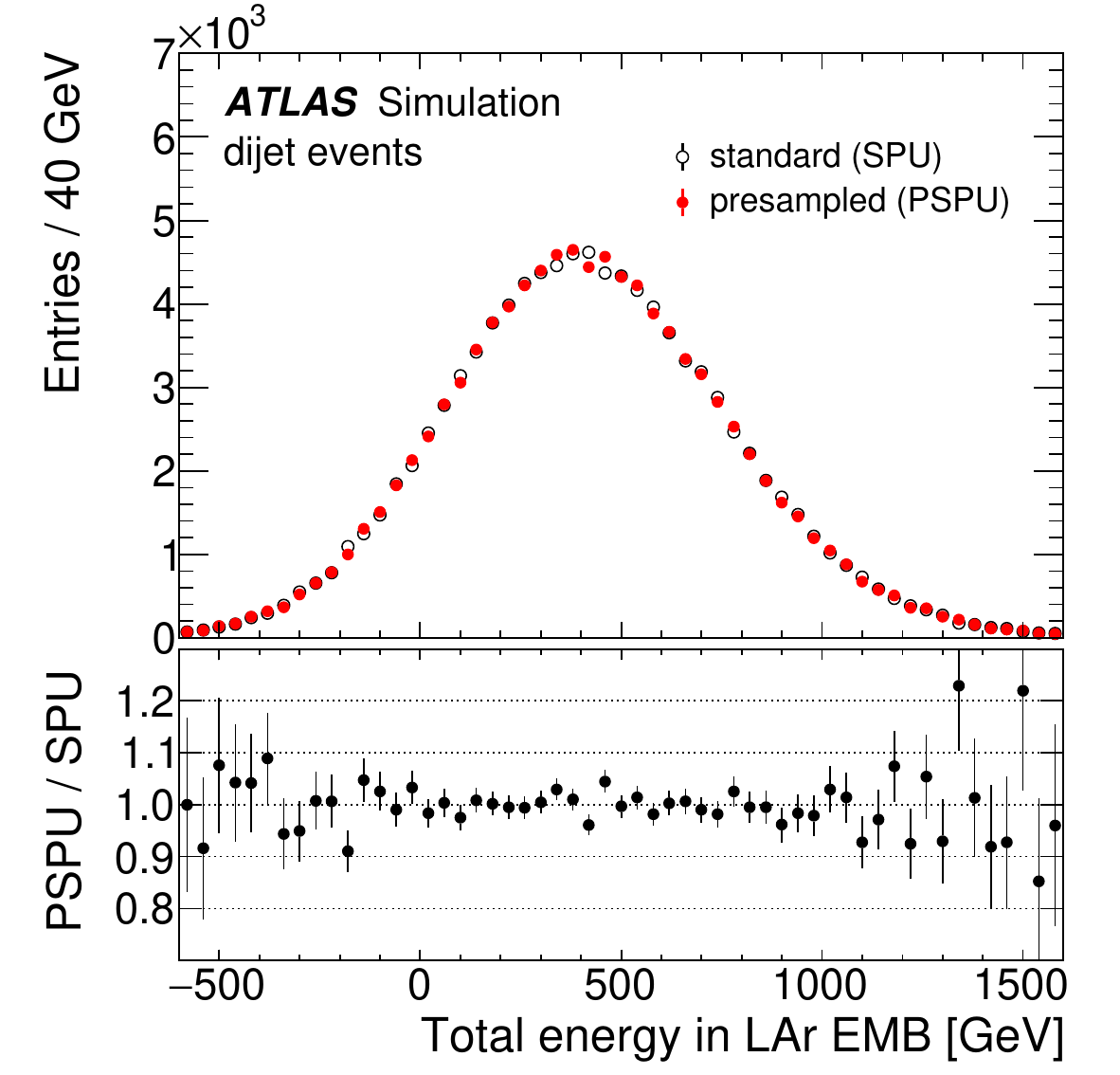}\label{fig:energy_EMB}}
\subfloat[]{\includegraphics[width=0.45\textwidth]{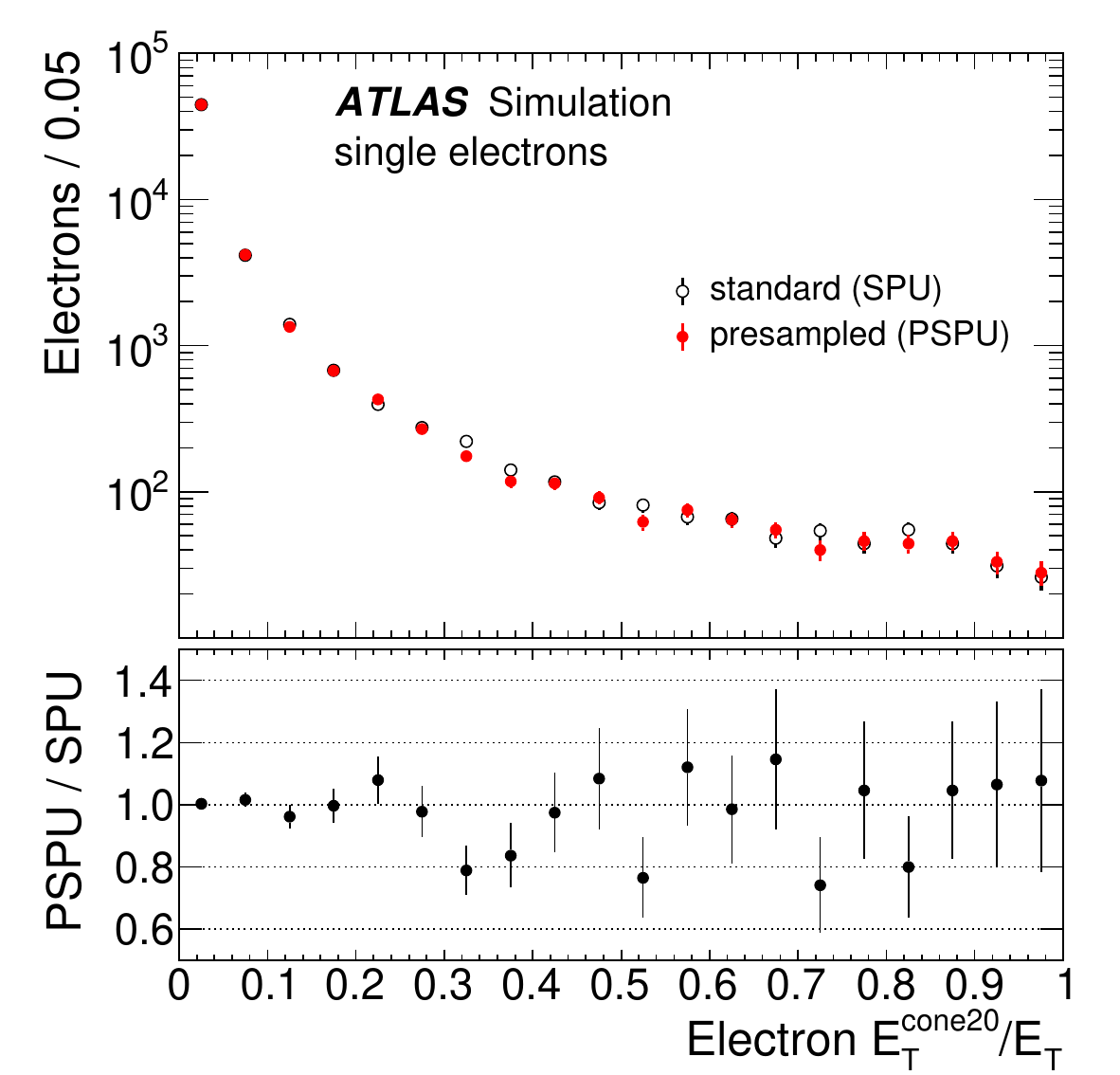}\label{fig:isolation}}\\
\subfloat[]{\includegraphics[width=0.45\textwidth]{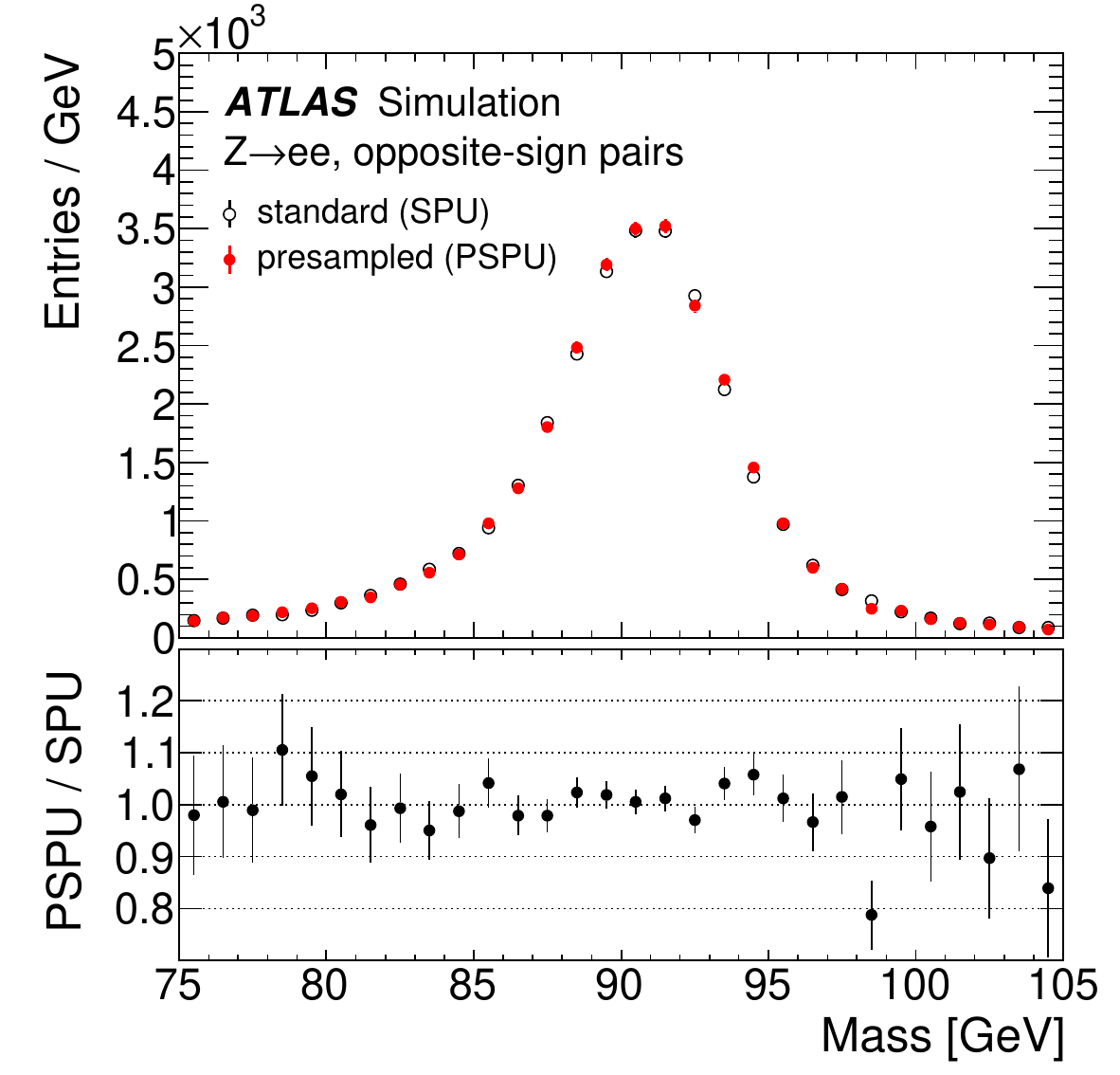}\label{fig:mee}}
\caption{A comparison between the standard digitisation (open black circles) and the
\mcoverlay (red filled circles) for (a) the total deposited energy distribution in the
electromagnetic barrel of the liquid-argon calorimeter in simulated dijet events, (b) the electron
isolation \( E^{\mathrm{cone20}}_{\mathrm{T}}/E_{\mathrm{T}} \) distribution for single electrons,
and (c) the opposite-sign electron-pair invariant mass distribution from simulated \( Z \rightarrow
e^+e^- \) events. The normalisation of the figures is arbitrary as it is simply proportional to the
number of events in the MC sample. The bottom panels show the ratios of the two distributions.}
\end{figure}

\Fig{\ref{fig:energy_EMB}} shows a comparison of the total energy deposited in the EMB
calorimeter by dijet events for the \mcoverlay and standard digitisation methods. This distribution
is sensitive to electronics and \pileup noise and shows that the simulation of the noise in the two
methods is similar.  \Fig{\ref{fig:isolation}} shows the distribution of a calorimeter isolation
quantity \( E^{\mathrm{cone20}}_{\mathrm{T}}/E_{\mathrm{T}} \) for simulated single-electron events.
This variable is calculated from topological clusters~\cite{PERF-2014-07} of energy deposits by
summing the transverse energies of such clusters within a cone of size \( \Delta R=0.2 \) around (but not including)
the candidate electron cluster.  It is sensitive to \pileup energy deposits close to the signal
electrons and is again similar for the two methods. \Fig{\ref{fig:mee}} shows the invariant
mass distribution of electron--positron pairs from simulated \( Z \rightarrow e^+e^- \) events. This
comparison shows that the energy scale and resolution of electrons from signal events agree for
the two methods.
 
\begin{figure}[htb!]
\centering
\subfloat[]{\includegraphics[width=0.45\textwidth]{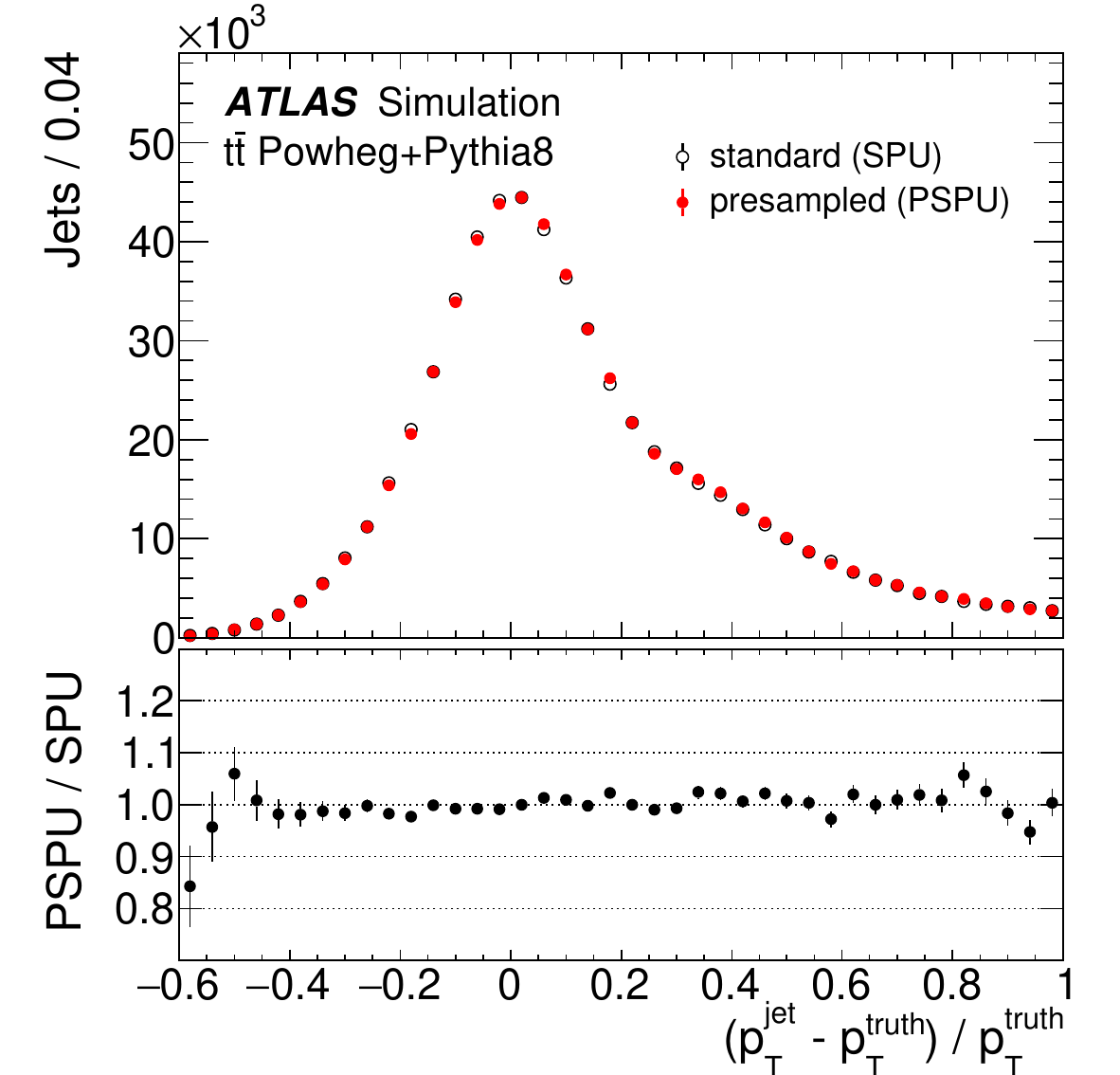}}
\subfloat[]{\includegraphics[width=0.45\textwidth]{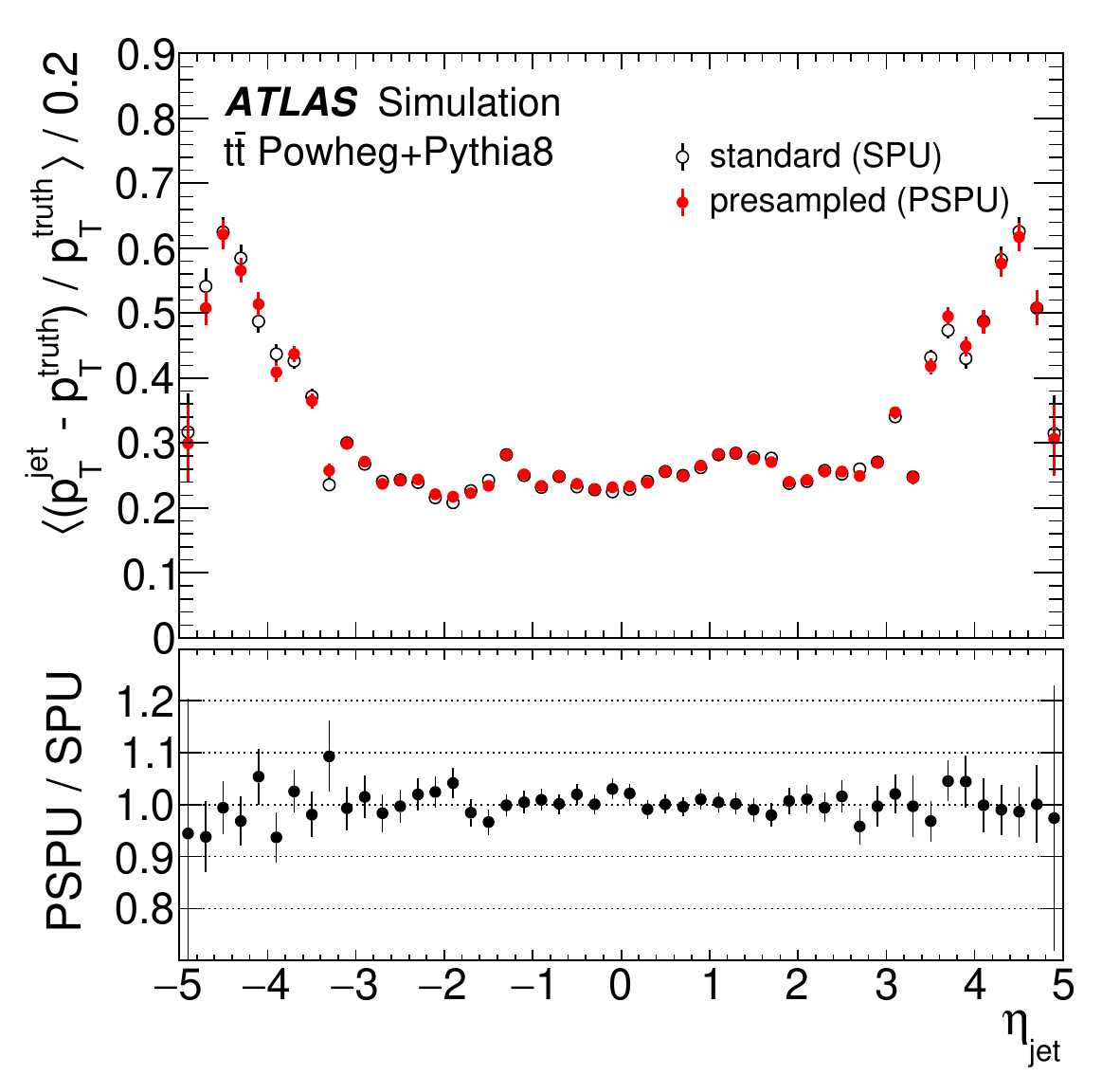}}\\
\subfloat[]{\includegraphics[width=0.45\textwidth]{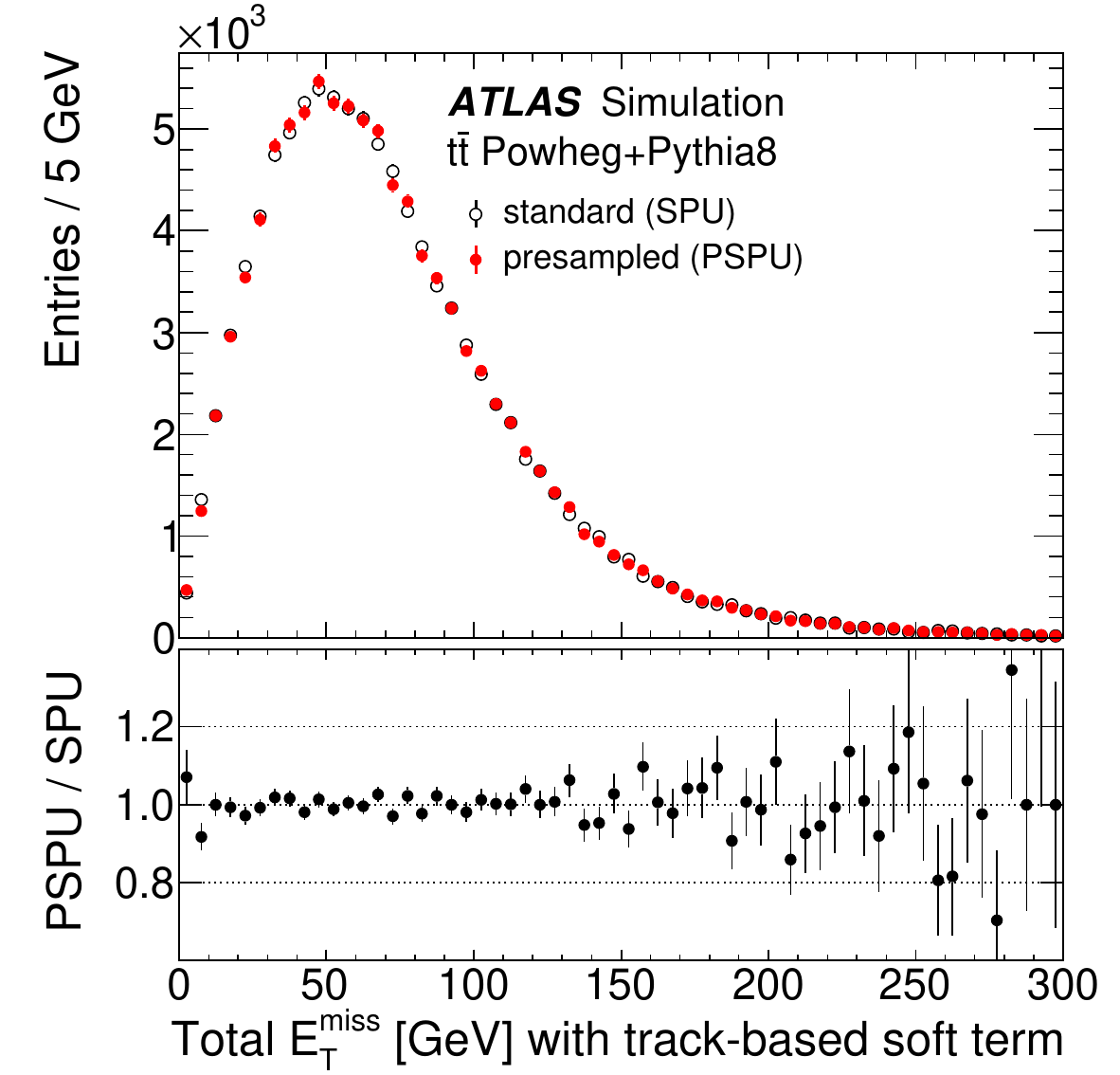}}
\subfloat[]{\includegraphics[width=0.45\textwidth]{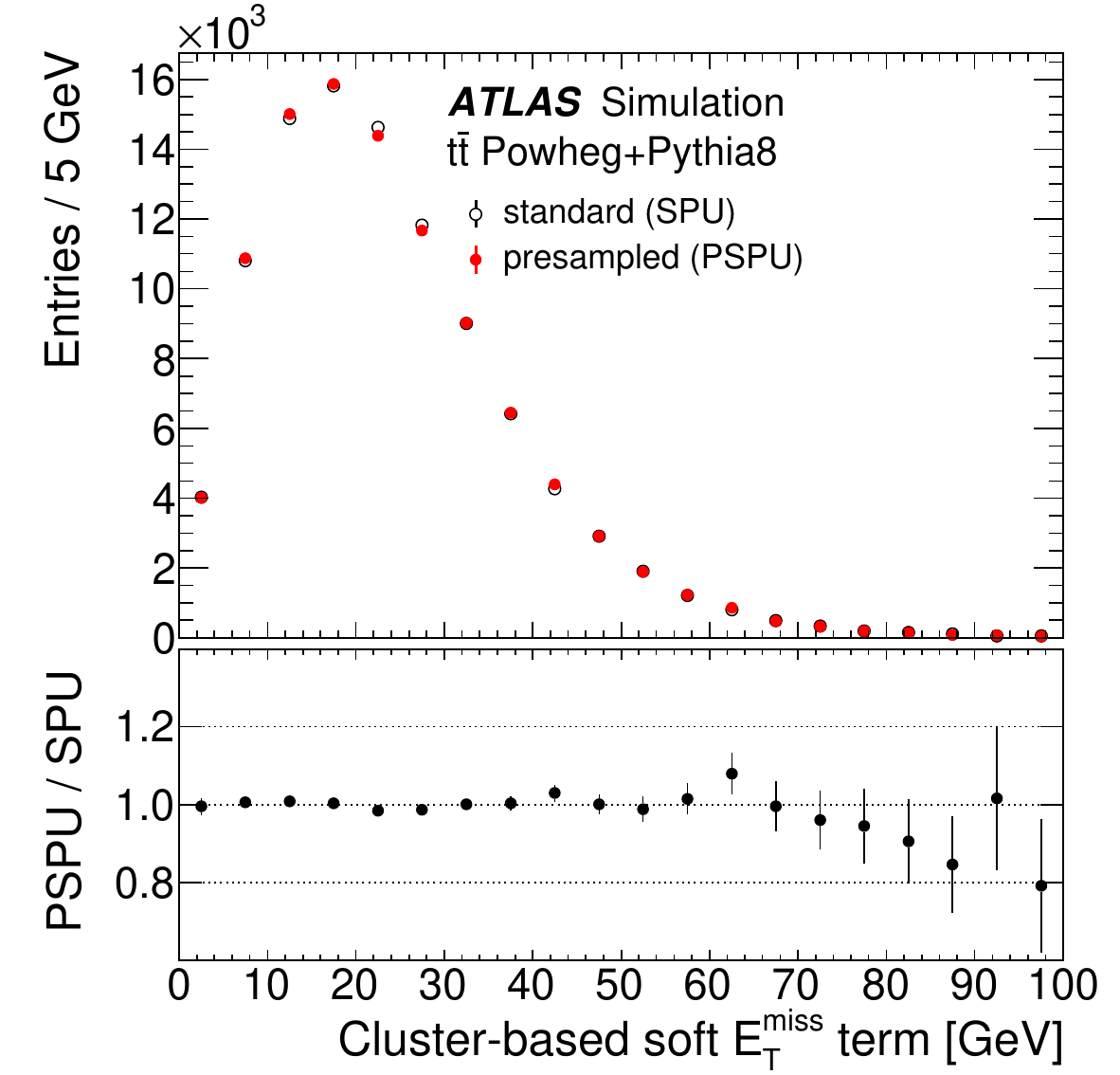}}
 
\caption{A comparison between the standard digitisation (open black circles)
and the \mcoverlay (red filled circles) in simulated \ttbar events for (a)
the jet \pt response, (b) the mean jet \pt response as a function of jet
pseudorapidity \(\eta_\textrm{jet}\), (c) the total \( \met \) distribution and
(d) the component of the \( \met \) from energy clusters in the calorimeter that are not associated with calibrated physics objects, known as the \textit{soft term}.
The bottom panels show the ratios of the two distributions.}
\label{fig:jetPt_response_all}
\end{figure}

\Fig{\ref{fig:jetPt_response_all}} shows the jet response in \ttbar MC events.  The jet \pt
is calibrated using a multi-stage procedure~\cite{PERF-2016-04} that accounts for several effects,
including \pileup. The \pileup correction is performed at an early stage of the calibration
procedure and removes excess energy due to both in-time and out-of-time \pileup. It is therefore
sensitive to the details of the \pileup emulation.
The shape of the
distribution (which is sensitive to noise modelling) and the average response versus \( \eta \) over the full
calorimeter acceptance are in good agreement for the two methods. Also shown in
\Fig{\ref{fig:jetPt_response_all}} is the distribution of missing transverse momentum \( \met \) for events in the same \ttbar
sample. The \textit{soft term} component, as reconstructed in the calorimeter, which is particularly sensitive to \pileup~\cite{PERF-2016-07} is shown as well.
Again, good agreement is observed for the two methods.
 
\section{Muon spectrometer}
\label{sec:muons}
The MS consists of four subdetectors: two providing high-precision tracking measurements and two
primarily providing trigger information.  The technologies used in these are different and, as with the ID, they require specific digitisation treatments for the \mcoverlay.  The main difference in the case of
the MS compared to the ID is that the occupancy is much lower.  This means that, while there is the
potential for loss of information in the \mcoverlay method if two sub-threshold hits occur in the
same detector channel, the probability of this occurring is much lower and the resulting effect is
found to be negligible.
 
\subsection{Detector readout and overlay procedure}
 
\paragraph{Monitored drift tubes (MDT):}  The MDTs consist of layers of drift tubes which are
designed to have a position resolution below \SI{80}{\micro\metre} per tube. If a particle traverses a drift
tube, ionisation is created and electrons drift to the anode wire. If the charge at that wire
exceeds a set threshold, the charge and the time are recorded, and both are converted to digital
information. For the \mcoverlay, the digital signals from the hard-scatter and \pileup events
are combined as follows. If a signal in a given tube is only present in either the hard-scatter event or
the \pileup event, that signal is copied to the output RDO.\@ If a signal is present in both, then
the two signal amplitudes are added, and the timing is taken to be the earlier of the two events.
 
\paragraph{Cathode strip chambers (CSC):} The CSCs are multiwire proportional chambers with
cathode strip readout which, by charge interpolation, provide a spatial resolution of \SI{60}{\micro\metre} in
the radial, or bending, plane and \SI{5}{mm} in the transverse, or \( \phi \), plane.  By combining the hits of a
track crossing all four chambers, a time resolution of \SI{4}{ns} is achieved, sufficient to identify the
bunch crossing. For each wire, the charge information per strip is recorded, then digitised and
stored in four time slices, each of \SI{50}{ns}. For the \mcoverlay, the charge deposited in each strip in the four
time slices is read out for the hard-scatter event and the \pileup event; the two signals are
then added separately per time slice and strip, taking care to ensure that the pedestal is subtracted
appropriately. The combined RDO resulting from these summed signals is then written out.

\paragraph{Resistive plate chambers (RPC):} The RPC system covers the region \( \abseta < 1.05 \) and is
composed of gaseous parallel-plate detectors. The position resolution is about \SI{1}{cm} in both the transverse and
longitudinal directions, and the time resolution is \SI{1.5}{ns}. If a muon crosses the \SI{2}{mm} space
between the two parallel resistive plates, an avalanche forms along the ionising track towards the
anode.  The signal is then read out via metallic strips mounted on the outer faces of the resistive
plates if it exceeds a given threshold; the time of the signal is also recorded.  For the \mcoverlay
the only relevant information is the time and the overlay is performed by taking, for each channel,
the earliest signal time between the hard-scatter and the \pileup events.
 
\paragraph{Thin gap chambers (TGC):} The TGCs cover the region \( 1.05< \abseta < 2.4 \). They have a typical
position resolution of 3--\SI{7}{mm} in the bending direction and 2--\SI{6}{mm} in the transverse direction, and a
time resolution of \SI{4}{ns}. The radial coordinate is measured by reading which TGC wire-group is hit;
the azimuthal coordinate is measured by reading which radial strip is hit.  For each wire, the time
at which a signal is above threshold is recorded and digitised and then written in the digit
format. As in the RPCs, the hard-scatter and \pileup events are combined by taking the earliest
arrival time of any hard-scatter or \pileup signal for a given wire.

\subsection{Validation results}
 
\begin{figure}[htb]
\centering
\subfloat[]{\includegraphics[width=0.45\textwidth]{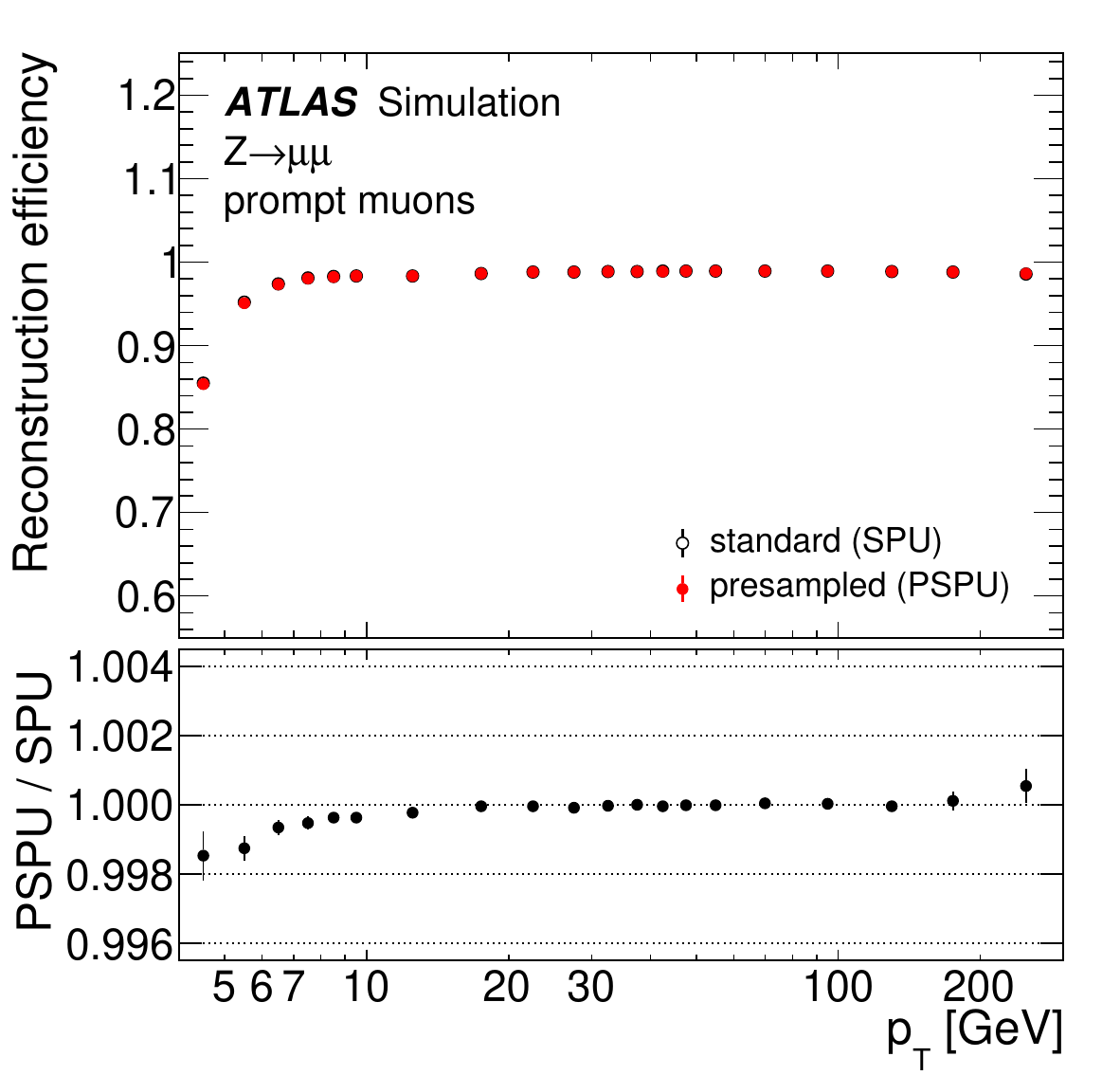}}
\subfloat[]{\includegraphics[width=0.45\textwidth]{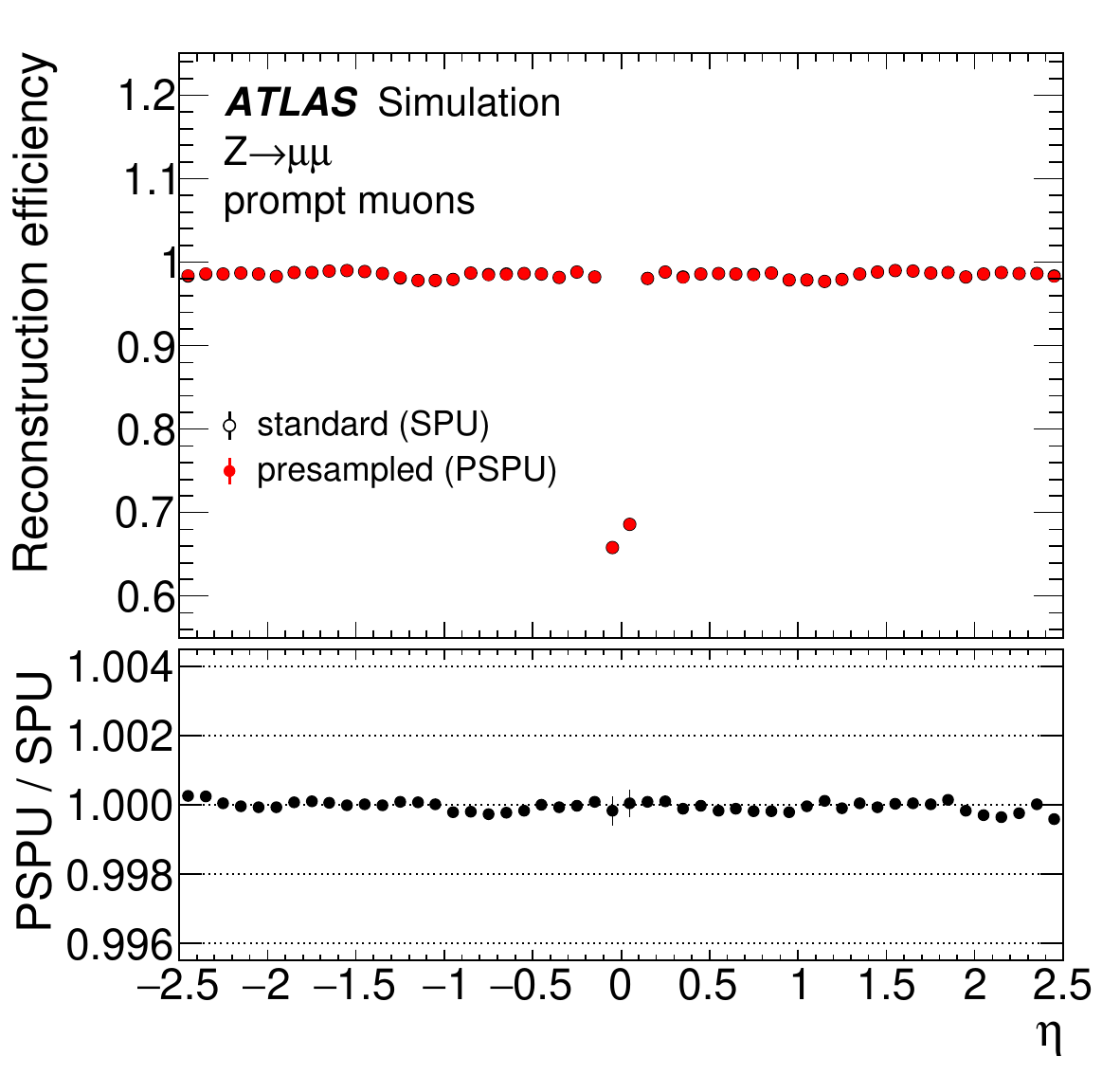}}\\
\subfloat[]{\includegraphics[width=0.45\textwidth]{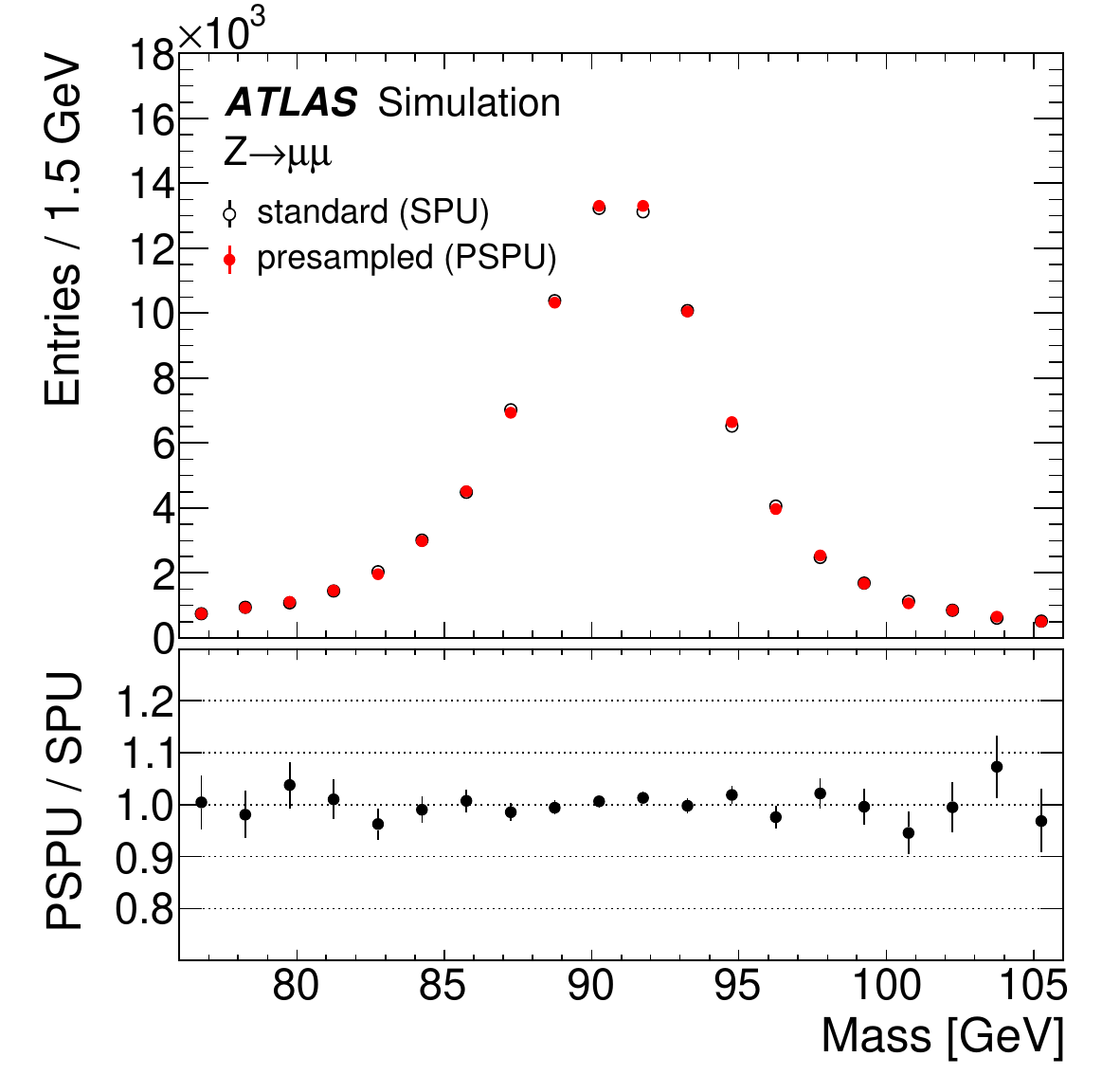}\label{fig:muon_reco_mass}}
\caption{The muon reconstruction efficiency versus (a) \pt and (b) \( \eta \) and (c) the dimuon invariant mass in simulated \( Z\to\mu^+\mu^- \) events.
The open black circles correspond to the standard digitisation and the red filled circles to \mcoverlay.
The bottom panels show the ratios of the corresponding distributions.}
\label{fig:muon_reco}
\end{figure}

The \mcoverlay procedure is validated by using muons from simulated \( Z \rightarrow \mu^+\mu^-\)
events and comparing their characteristics with those after the standard \pileup digitisation
procedure. \Fig{\ref{fig:muon_reco}} shows the reconstruction efficiency of muons as a function
of \pt and \( \eta \) for the two methods.  They agree to better than 0.1\% for nearly the
entire \pt and \( \eta \) range. \Fig{\ref{fig:muon_reco_mass}} shows the invariant mass of the two muons
for the same event sample. Also here, good agreement is observed between the two methods.
 
\section{Trigger}
\label{sec:trig}
The L1 trigger receives inputs from the L1 calorimeter (L1Calo) and L1 muon triggers. The L1Calo
decision is formed using reduced granularity inputs from the LAr and Tile calorimeters. The L1
muon trigger receives signals from the RPCs in the barrel and from the TGCs in the endcaps as is described
in \Sect{\ref{sec:muons}}. After the L1 trigger decision, the HLT has access to the data from the
full detector to perform a refined analysis. The trigger decisions and all reconstructed objects are
stored in a dedicated record of the accepted event.
 
The L1 hardware trigger is simulated using dedicated algorithms that strive to perform a bit-wise
correct emulation of the trigger decision including any trigger objects that the hardware produces.
The HLT runs on the output of the L1 trigger using the same simulation software as used for data.
The following sections discuss the L1 calorimeter trigger and the overall HLT performance. No dedicated
changes were required to the muon trigger simulation beyond what is discussed for the general
simulation in \Sect{\ref{sec:muons}}.  While the HLT software itself remains unchanged between the two methods, it depends on the inputs form the various subdetectors that do differ and hence serves as an additional validation.
 
\subsection{L1 calorimeter trigger simulation}
The inputs to the L1Calo trigger processors are \textit{trigger towers}~\cite{TRIG-2016-01}. These are
formed in the on-detector electronics by summation of the analogue voltage pulses from calorimeter
cells in groups of \(\Delta\eta \times \Delta \phi \sim 0.1 \times \pi/32\), separately in the
electromagnetic and hadronic calorimeter systems. These signals are then transmitted over \SI{70}{m} long,
individually shielded, twisted-pair cables to the trigger electronics, where they are digitised with
a least-count equivalent to \SI{250}{\MeV} transverse energy and a sampling frequency of
\SI{40}{MHz}. A custom digital processor, consisting of filters, comparators and look-up tables,
analyses the shape of the digitised pulse in order to identify which bunch crossing it came from.  It
also corrects for shifts in the baseline caused by the interplay of in-time and out-of-time
pile-up due to the LHC bunch structure, subtracts pedestals and applies noise cuts.  Finally, it
provides calibrated transverse energy \ET values for use in the trigger algorithms on scales of
\SI{500}{\MeV/count} for the electron, photon and \( \tau \)-lepton triggers and \SI{1}{\GeV/count} for all
other triggers.
 
In the simulation, the analogue signals received from the calorimeters are represented by objects
containing a vector of floating-point values, corresponding to the amplitudes of the pulses sampled
at \SI{25}{ns} intervals. These are then quantised, with the addition of noise from the digitisation
system, and passed through a precise simulation of the signal processing performed by the
trigger electronics.  The calorimeter objects are formed from calorimeter hits, using a model of the
pulse shaping and the noise from the readout and summation chain.
 
For \mcoverlay, the \textit{analogue} calorimeter objects are merged before the trigger
digitisation and processing are performed. This then allows the unmodified trigger simulation to be performed on
the merged data, and it avoids any possible bias due to merging data that have been quantised on a
relatively coarse scale. The merging is performed by an additional algorithm, which is run during
the \pileup merging prior to the trigger simulation to create a set of merged calorimeter towers.
The merging itself uses the calorimeter object identifiers to match corresponding towers in the
hard-scatter and \pileup event collections, and the amplitudes of the signals of the same towers in both
events are summed. A new collection of objects containing the summed amplitudes is then created and
written to the output stream.
 
\begin{figure}[htb]
\centering
\begin{tabular}{cc}
\subfloat[]{\includegraphics[width=0.45\textwidth]{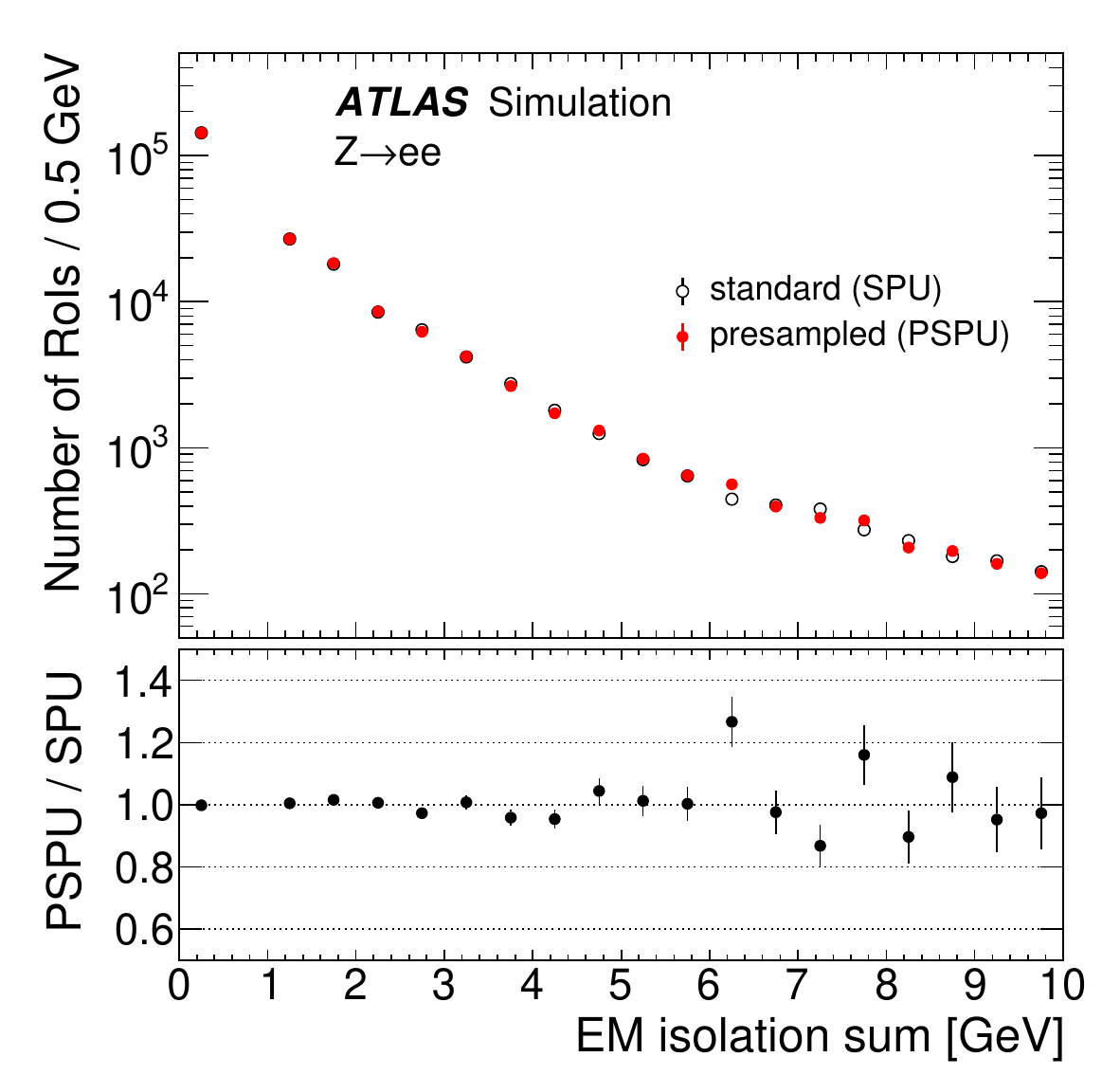}} &
\subfloat[]{\includegraphics[width=0.45\textwidth]{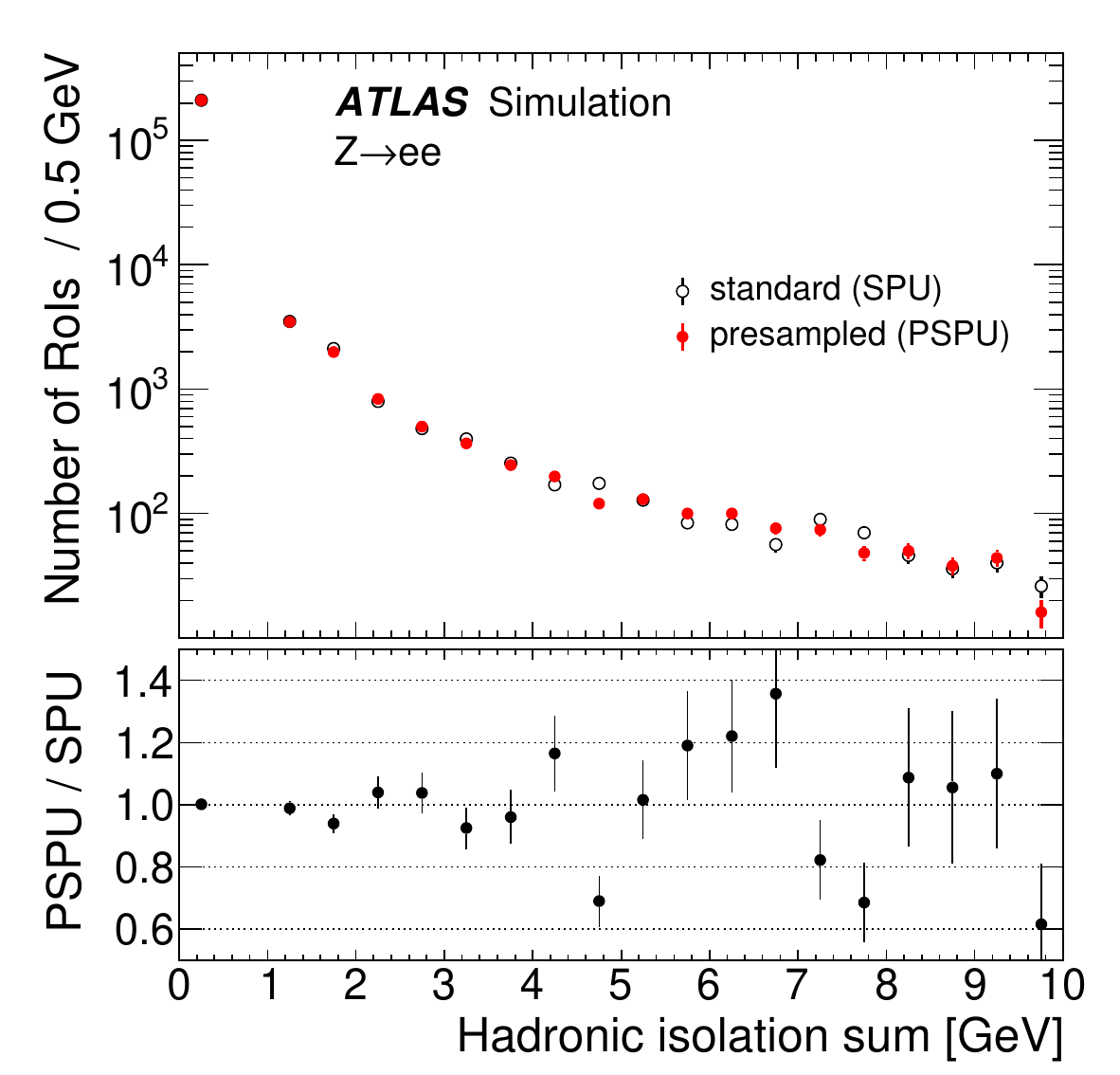}} \\
\end{tabular}
\caption{Distributions of \ET in the isolation regions of the L1Calo \( e/\gamma \) trigger, (a) in the electromagnetic calorimeter and (b) in the hadronic calorimeter.
The standard digitisation (black open circles) is compared with the \mcoverlay (red filled circles). The distributions are for regions around electrons in \(Z \rightarrow e^+ e^-\) events, which
are dominated by electronic noise and \pileup.
The bottom panels show the ratios of the two distributions.
}
\label{fig:L1Calo_Isolation}
\end{figure}

\Fig{\ref{fig:L1Calo_Isolation}} shows the L1Calo \ET distributions in isolation regions around electrons
in \(Z \rightarrow e^+ e^-\) events, which are sensitive to the \pileup \ET deposits close
to the electrons. Good agreement is seen between the \standardandover \simcs.
 
\subsection{HLT simulation and performance}
 
After being accepted by the L1 trigger, the events are processed by the HLT using finer-granularity calorimeter
information, precision measurements from the muon system and tracking information from the inner
detector. As needed, the HLT reconstruction can be executed either for the full event or within
smaller, isolated regions of interest (RoIs) identified by the L1 trigger. In order to reduce the
processing time, most HLT triggers use a two-stage approach with a fast (trigger-specific)
first-pass reconstruction to reject the majority of events and a slower, higher-precision (offline-like)
reconstruction for the remaining events.

\begin{figure}[htb]
\centering
\subfloat[]{\includegraphics[width=0.5\textwidth]{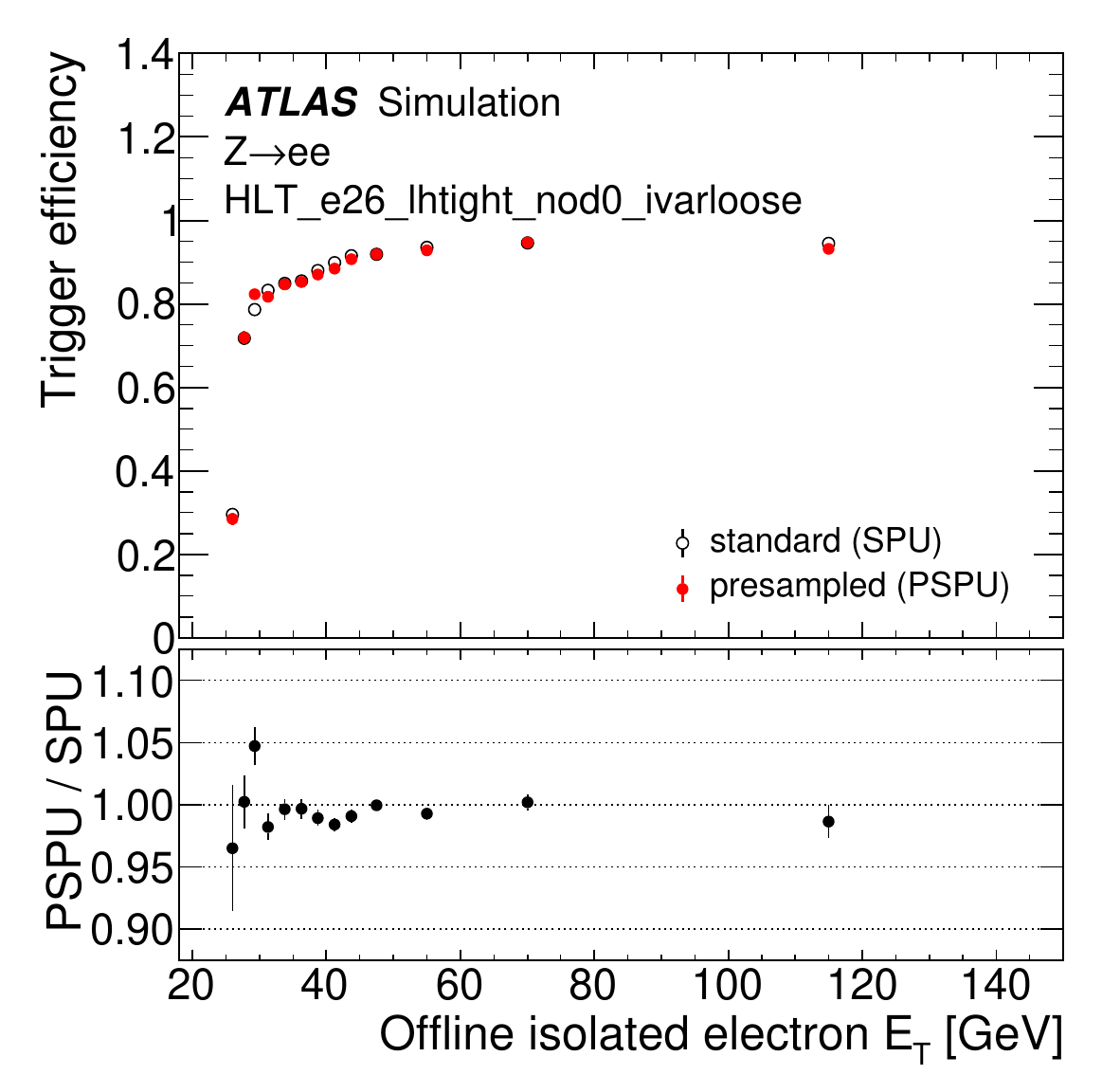}}
\subfloat[]{\includegraphics[width=0.5\textwidth]{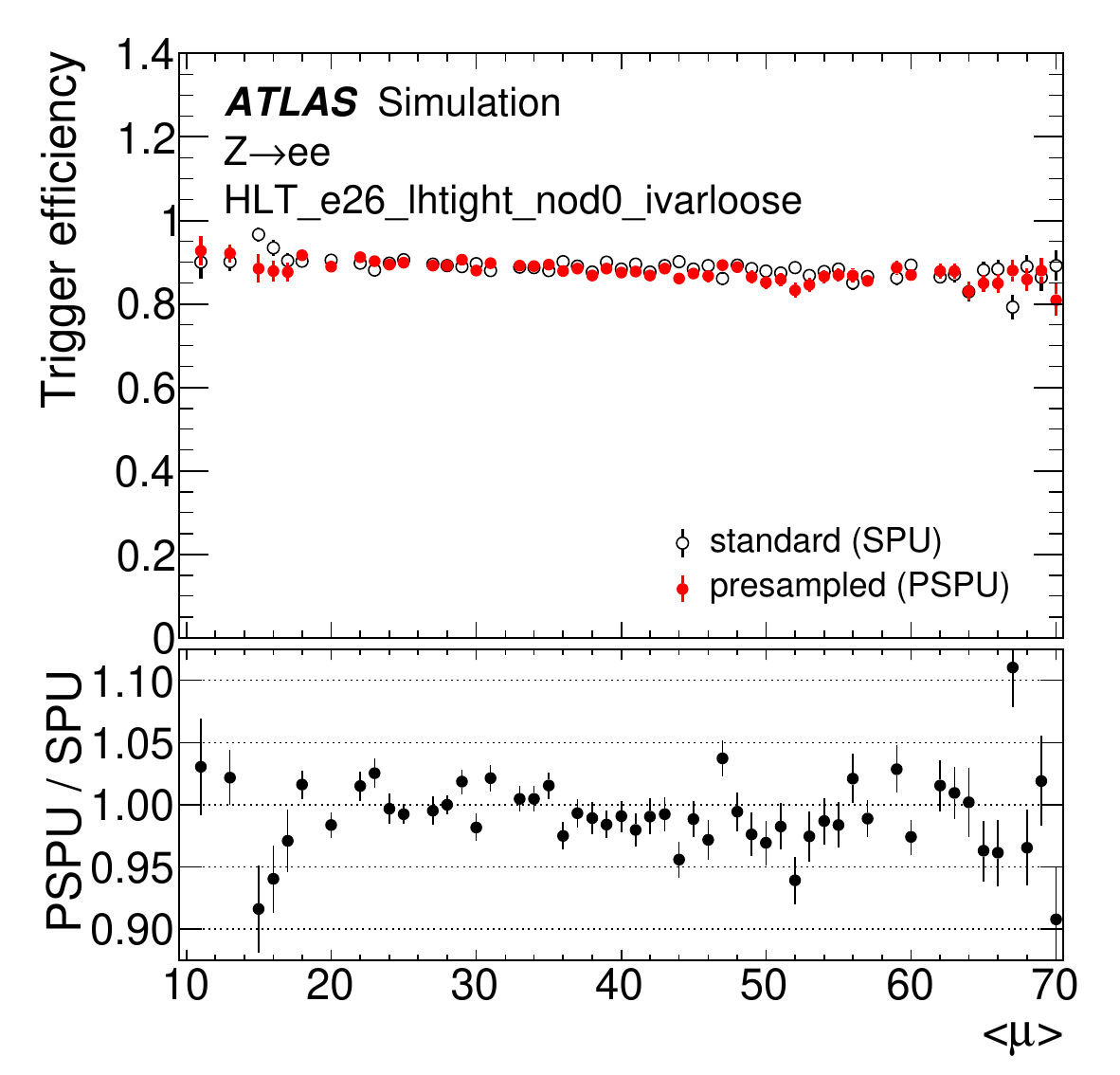}}
\caption{The combined L1 and HLT trigger efficiency of the \SI{28}{\GeV} electron trigger from simulated \( Z \to e^+e^- \) events (red filled circles)
as a function of (a) \ET and (b) \pileup for the standard digitisation (open black circles) and \mcoverlay (red filled circles). The bottom panels show the ratios of the two distributions.
}
\label{fig:trig_egamma_eff}
\end{figure}

\begin{figure}[htb]
\centering
\subfloat[]{\includegraphics[width=0.5\textwidth]{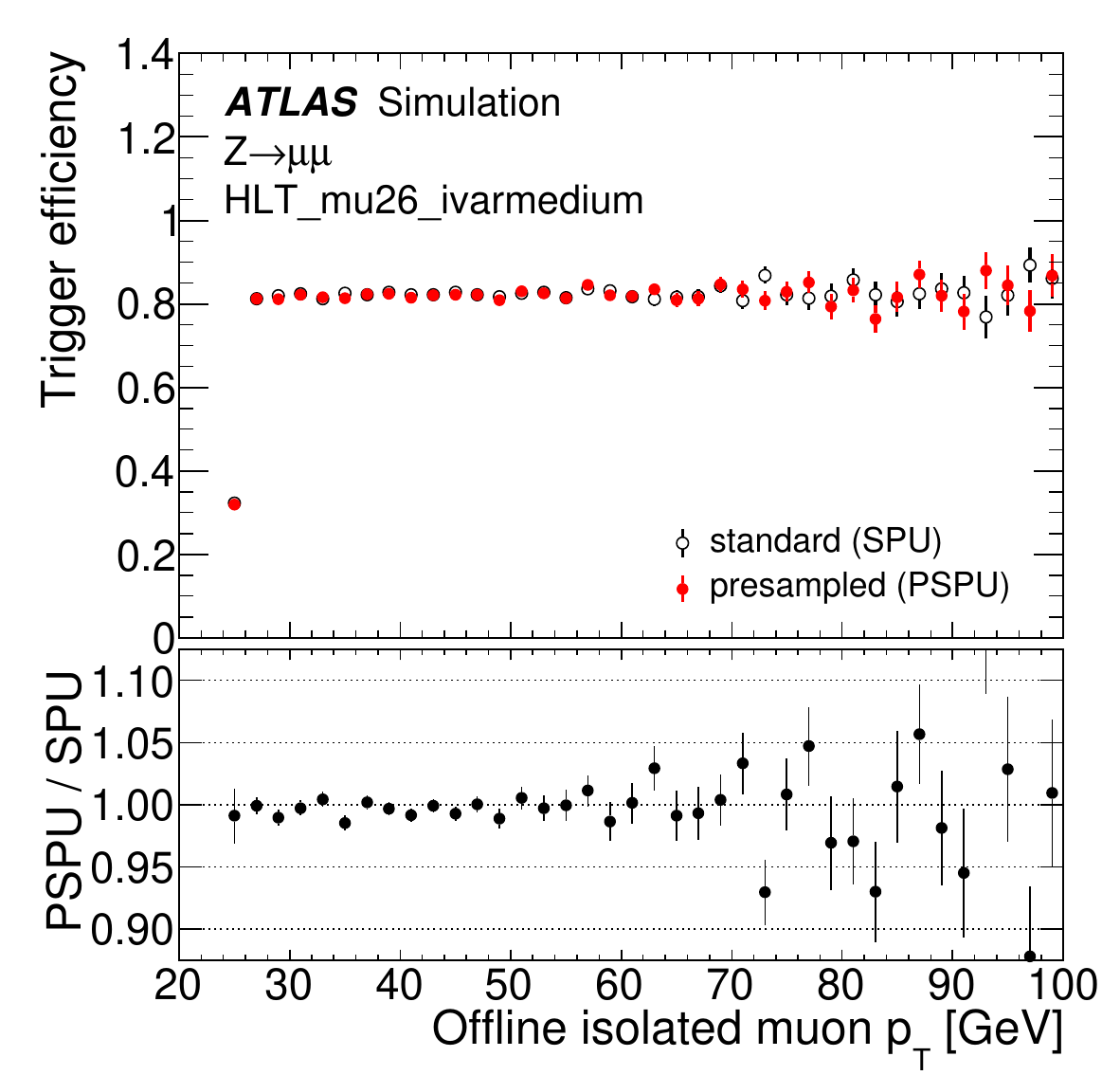}}
\subfloat[]{\includegraphics[width=0.5\textwidth]{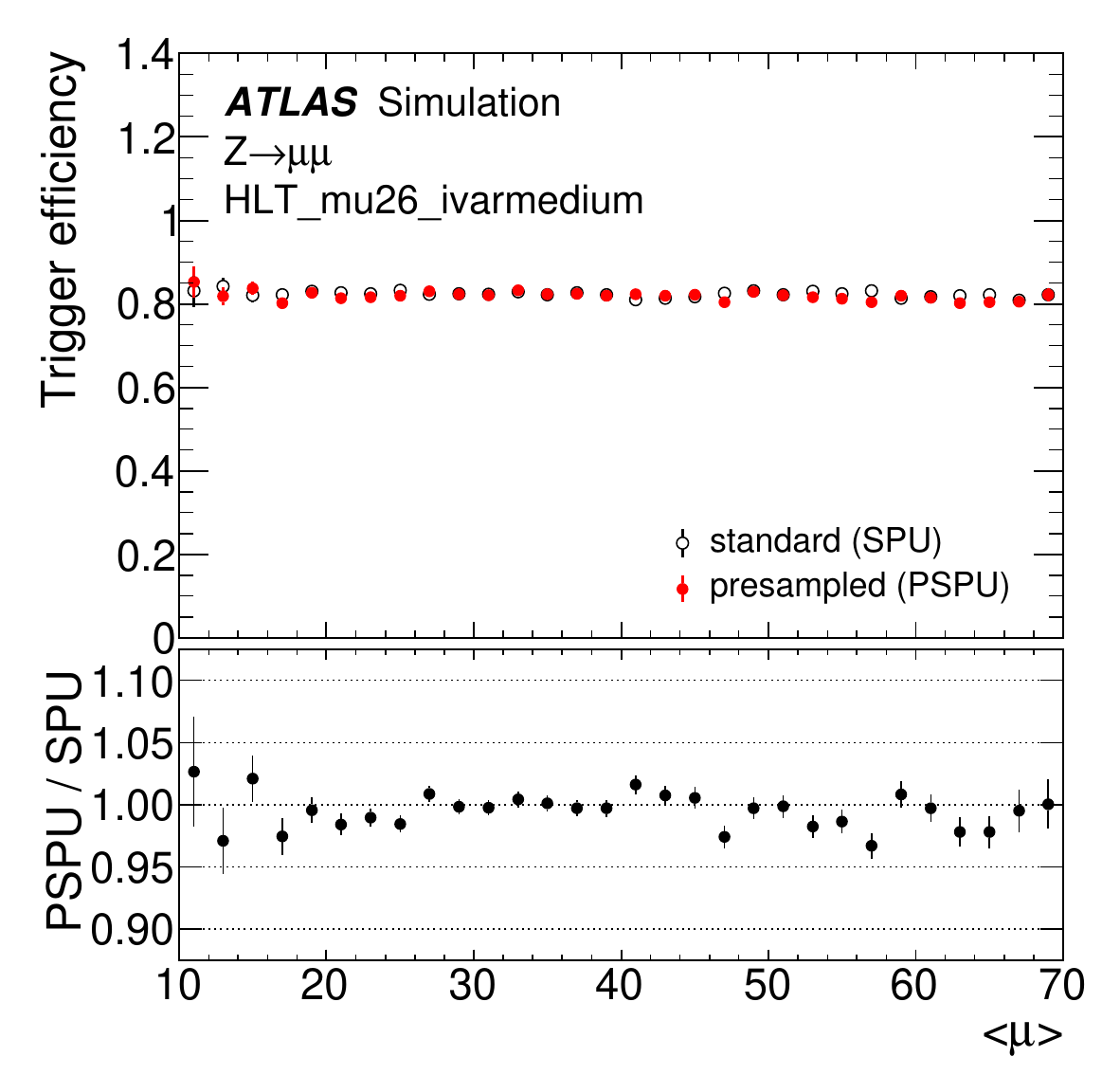}}
\caption{The combined L1 and HLT trigger efficiency of the \SI{26}{\GeV} muon trigger from simulated \( Z \to \mu^+\mu^- \) events
as a function of (a) \pt and (b) \pileup for the standard digitisation (open black circles) and \mcoverlay
(red filled circles). The bottom panels show the ratios of the two distributions.
}
\label{fig:trig_muon_eff}
\end{figure}

The reconstruction of electron (muon) candidates requires the matching of a calorimeter cluster
(muon spectrometer track) to a track in the inner detector and is therefore sensitive to changes in
the inner detector, calorimeter and muon spectrometer
reconstruction. \Fig{\ref{fig:trig_egamma_eff}} shows the trigger efficiency of the primary
\SI{28}{\GeV} electron trigger measured with simulated \( Z \to e^+e^- \) events for the \standardandover \simcs. Similarly, \Fig{\ref{fig:trig_muon_eff}} shows the trigger efficiency of the
primary \SI{26}{\GeV} muon trigger measured with simulated \( Z \to \mu^+\mu^- \) events. No significant
differences are observed in the trigger efficiency between the \mcoverlayandstd \simcs.
 
\begin{figure}[htb]
\centering
\subfloat[]{\includegraphics[width=0.5\textwidth]{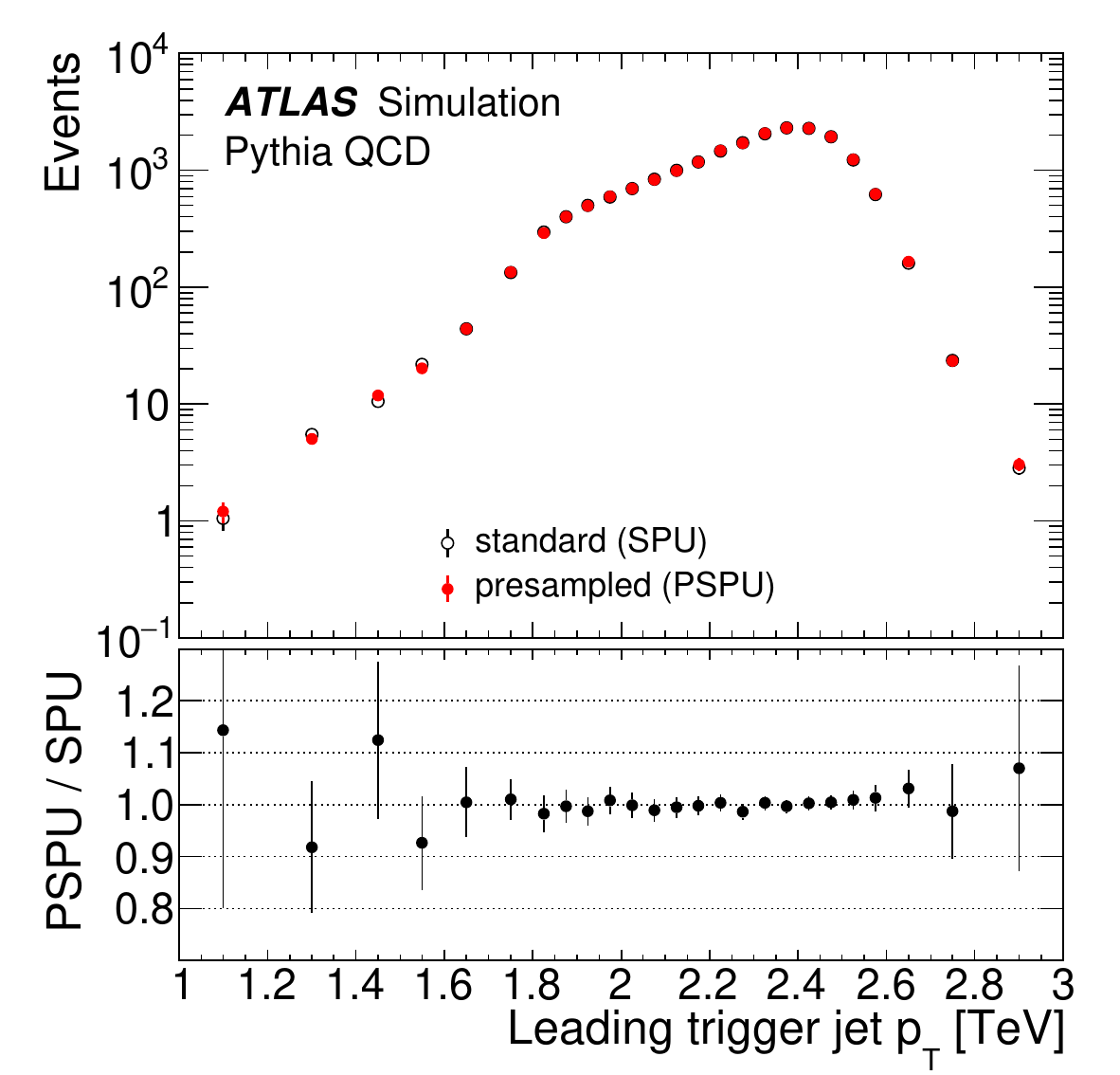}}
\subfloat[]{\includegraphics[width=0.5\textwidth]{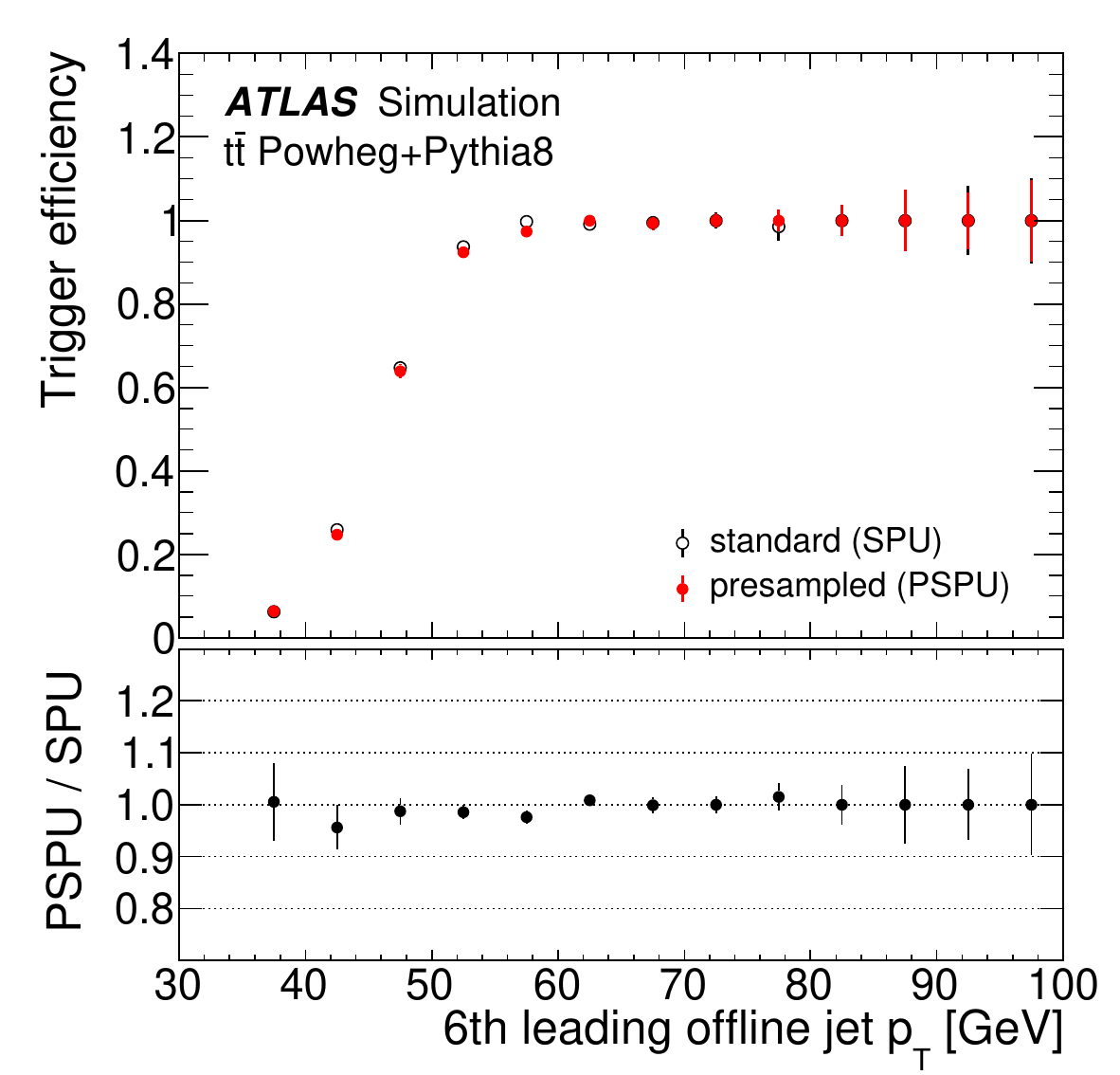}}
\caption{(a) The \pt of the leading jet in all events with a triggered jet and (b) the trigger efficiency
of the 6-jet trigger requiring a jet \( \pt>\SI{45}{\GeV} \) as a function of the \pt of the sixth leading jet
for the standard digitisation (open black circles) and \mcoverlay (red filled circles). The bottom panels show the ratios of the two distributions.
}
\label{fig:trig_hlt_jet}
\end{figure}

Jet and \met triggers are mainly based on the calorimeter reconstruction and are especially
sensitive to changes in the simulation of low-\pt jets. \Fig{\ref{fig:trig_hlt_jet}} shows the
\pt distribution of the leading jet and the trigger efficiency as a
function of the sixth leading jet \pt for a multi-jet trigger requiring six jets with a \pt larger than
\SI{45}{\GeV}. Good agreement between the \standardandover \simcs is observed in both cases.
 
All other triggers relevant to the ATLAS physics programme were also studied and no notable differences between the two methods were observed.

\FloatBarrier
\section{Conclusions}
A new method for reproducing the impact of \pileup interactions on the ATLAS detector performance is presented, based on overlaying presampled \pileup events on the hard-scatter event of interest during the digitisation.   The method is validated separately for each ATLAS detector system and the trigger. In all cases, it is possible to achieve good agreement with the \standardmethod \simc which has been used up to now. For a large variety of quantities, detailed comparisons are made between the two methods, and all the differences are found to be small\cbg{Can we quantify?  Will depend how updated plots from Tadej look with new errors}, so that the impact on physics analyses is considered negligible.
 
The \mcoverlay method is shown to use significantly less computing resources than the standard method used so far within ATLAS.\@
For the \RunTwo \pileup distribution and software, the CPU resources required for the entire MC \simc are reduced by around 20\%.

\section*{Acknowledgements}
 

We thank CERN for the very successful operation of the LHC, as well as the
support staff from our institutions without whom ATLAS could not be
operated efficiently.
 
We acknowledge the support of ANPCyT, Argentina; YerPhI, Armenia; ARC, Australia; BMWFW and FWF, Austria; ANAS, Azerbaijan; SSTC, Belarus; CNPq and FAPESP, Brazil; NSERC, NRC and CFI, Canada; CERN; ANID, Chile; CAS, MOST and NSFC, China; Minciencias, Colombia; MSMT CR, MPO CR and VSC CR, Czech Republic; DNRF and DNSRC, Denmark; IN2P3-CNRS and CEA-DRF/IRFU, France; SRNSFG, Georgia; BMBF, HGF and MPG, Germany; GSRT, Greece; RGC and Hong Kong SAR, China; ISF and Benoziyo Center, Israel; INFN, Italy; MEXT and JSPS, Japan; CNRST, Morocco; NWO, Netherlands; RCN, Norway; MNiSW and NCN, Poland; FCT, Portugal; MNE/IFA, Romania; JINR; MES of Russia and NRC KI, Russian Federation; MESTD, Serbia; MSSR, Slovakia; ARRS and MIZ\v{S}, Slovenia; DST/NRF, South Africa; MICINN, Spain; SRC and Wallenberg Foundation, Sweden; SERI, SNSF and Cantons of Bern and Geneva, Switzerland; MOST, Taiwan; TAEK, Turkey; STFC, United Kingdom; DOE and NSF, United States of America. In addition, individual groups and members have received support from BCKDF, CANARIE, Compute Canada, CRC and IVADO, Canada; Beijing Municipal Science \& Technology Commission, China; COST, ERC, ERDF, Horizon 2020 and Marie Sk{\l}odowska-Curie Actions, European Union; Investissements d'Avenir Labex, Investissements d'Avenir Idex and ANR, France; DFG and AvH Foundation, Germany; Herakleitos, Thales and Aristeia programmes co-financed by EU-ESF and the Greek NSRF, Greece; BSF-NSF and GIF, Israel; La Caixa Banking Foundation, CERCA Programme Generalitat de Catalunya and PROMETEO and GenT Programmes Generalitat Valenciana, Spain; G\"{o}ran Gustafssons Stiftelse, Sweden; The Royal Society and Leverhulme Trust, United Kingdom.
 
The crucial computing support from all WLCG partners is acknowledged gratefully, in particular from CERN, the ATLAS Tier-1 facilities at TRIUMF (Canada), NDGF (Denmark, Norway, Sweden), CC-IN2P3 (France), KIT/GridKA (Germany), INFN-CNAF (Italy), NL-T1 (Netherlands), PIC (Spain), ASGC (Taiwan), RAL (UK) and BNL (USA), the Tier-2 facilities worldwide and large non-WLCG resource providers. Major contributors of computing resources are listed in Ref.~\cite{ATL-SOFT-PUB-2020-001}.
 

\printbibliography
 
\clearpage
 
\begin{flushleft}
\hypersetup{urlcolor=black}
{\Large The ATLAS Collaboration}

\bigskip

\AtlasOrcid[0000-0002-6665-4934]{G.~Aad}$^\textrm{\scriptsize 101}$,    
\AtlasOrcid[0000-0002-5888-2734]{B.~Abbott}$^\textrm{\scriptsize 127}$,    
\AtlasOrcid[0000-0002-7248-3203]{D.C.~Abbott}$^\textrm{\scriptsize 102}$,    
\AtlasOrcid[0000-0002-2788-3822]{A.~Abed~Abud}$^\textrm{\scriptsize 36}$,    
\AtlasOrcid[0000-0002-1002-1652]{K.~Abeling}$^\textrm{\scriptsize 53}$,    
\AtlasOrcid[0000-0002-2987-4006]{D.K.~Abhayasinghe}$^\textrm{\scriptsize 93}$,    
\AtlasOrcid[0000-0002-8496-9294]{S.H.~Abidi}$^\textrm{\scriptsize 29}$,    
\AtlasOrcid[0000-0002-8279-9324]{O.S.~AbouZeid}$^\textrm{\scriptsize 40}$,    
\AtlasOrcid{N.L.~Abraham}$^\textrm{\scriptsize 155}$,    
\AtlasOrcid[0000-0001-5329-6640]{H.~Abramowicz}$^\textrm{\scriptsize 160}$
% ...
\bigskip
\\

% ...
$^{29}$Physics Department, Brookhaven National Laboratory, Upton NY; United States of America.\\
% ...
$^{36}$CERN, Geneva; Switzerland.\\
% ...
$^{40}$Niels Bohr Institute, University of Copenhagen, Copenhagen; Denmark.\\
% ...
$^{53}$II. Physikalisches Institut, Georg-August-Universit\"{a}t G\"ottingen, G\"ottingen; Germany.\\
% ...
$^{93}$Department of Physics, Royal Holloway University of London, Egham; United Kingdom.\\
% ...
$^{101}$CPPM, Aix-Marseille Universit\'e, CNRS/IN2P3, Marseille; France.\\
$^{102}$Department of Physics, University of Massachusetts, Amherst MA; United States of America.\\
% ...
$^{127}$Homer L. Dodge Department of Physics and Astronomy, University of Oklahoma, Norman OK; United States of America.\\
% ...
$^{155}$Department of Physics and Astronomy, University of Sussex, Brighton; United Kingdom.\\
% ...
$^{160}$Raymond and Beverly Sackler School of Physics and Astronomy, Tel Aviv University, Tel Aviv; Israel.\\
% ...

\end{flushleft}


\end{document}